\documentclass[preprint,12pt]{elsarticle}

\usepackage{amssymb}
\usepackage{graphicx}
\usepackage[numbers]{natbib}
\usepackage{hyperref}
\usepackage{cleveref}
\usepackage{epstopdf}

\setcitestyle{numeric-comp}

\journal{ }

\newcommand{\kbar}{$\bar{\mathrm{K}}~$}
\newcommand{\kbarN}{$\bar{\mathrm{K}}$N }
\newcommand{\Lambdan}{$\mathrm{\Lambda}$}
\newcommand{\Sigman}{$\mathrm{\Sigma}$}
\newcommand{\Phin}{$\mathrm{\Phi}$}

\begin{document}

\begin{frontmatter}



\title{K$^-$ multi-nucleon absorption cross sections and branching ratios in \Lambdan p  and \Sigman$^0$p final states}


\author[addr2]{R. Del Grande}
\ead{raffaele.delgrande@lnf.infn.it}
\author[addr3,addr2]{K. Piscicchia}
\ead{kristian.piscicchia@gmail.com}
\author[addr4,addr5]{O.~Vazquez Doce}
\ead{oton.vd@cern.ch}
\author[addr6]{M.~Cargnelli}
\author[addr2]{C.~Curceanu}
\author[addr4,addr5]{L.~Fabbietti} 
\author[addr6]{J.~Marton}
\author[addr7]{P.~Moskal}
\author[addr8]{A.~Ramos}
\author[addr2]{A.~Scordo}
\author[addr2]{D.~Sirghi}
\author[addr7]{M.~Skurzok}
\author[addr10]{S.~Wycech}
\author[addr6]{J.~Zmeskal}
\author[addr7]{E.~Czerwinski}
\author[addr11]{V.~De~Leo}
\author[addr2]{P.~Fermani}
\author[addr12,addr13]{G.~Mandaglio}
\author[addr2,addr14]{M.~Martini}
\author[addr11]{N.~Raha}
\author[addr15,addr16]{A.~Selce}
\author[addr7]{M.~Silarski}

\address[addr2]{INFN - Laboratori Nazionali di Frascati, Via Enrico Fermi 40, 00044, Frascati, Italy}
\address[addr3]{CENTRO FERMI - Museo Storico della Fisica e Centro Studi e Ricerche ``Enrico Fermi'', 00184, Rome, Italy}
\address[addr4]{Excellence Cluster ``Origin and Structure of the Universe'', 85748 Garching, Germany}
\address[addr5]{Physik Department E12, Technische Universit{\"a}t M{\"u}nchen, 85748 Garching, Germany}
\address[addr6]{Stefan-Meyer-Institut f\"ur Subatomare Physik, 1090 Wien, Austria}
\address[addr7]{Institute of Physics, Jagiellonian University, 30-348 Cracow, Poland}
\address[addr8]{Departament de Fisica Quantica i Astrofisica and Institut de Ciencies del Cosmos, Universitat de Barcelona, Marti i Franques 1, 08028 Barcelona, Spain}
\address[addr10]{National Centre for Nuclear Research, 00681 Warsaw, Poland}
\address[addr11]{INFN Sezione di Roma Tor Vergata, Roma, Italy}
\address[addr12]{Dipartimento di Scienze Chimiche, Biologiche, Farmaceutiche ed Ambientali dell'Universit\`a di Messina, Messina, Italy}
\address[addr13]{INFN Sezione di Catania, Catania, Italy}
\address[addr14]{Dipartimento di Scienze e Tecnologie applicate, Universit\`a ``Guglielmo Marconi'', Roma, Italy}
\address[addr15]{Dipartimento di Matematica e Fisica dell'Universit\`a ``Roma Tre'', Roma, Italy}
\address[addr16]{INFN Sezione di Roma Tre, Roma, Italy}

\begin{abstract}

The determination of low-energy cross sections and branching ratios of the K$^-$ multi-nucleon absorption processes in \Lambdan p and \Sigman$^0$p final states performed by the AMADEUS collaboration is presented.
Low momentum K$^-$ ($p_\mathrm{K} \simeq$ 127 MeV/c) produced at the DA\Phin NE collider are impinged on a Carbon target within the KLOE detector and the two and three nucleon absorption processes are disentangled by comparing the experimental data to phenomenological calculations.  
The \Lambdan p spectra are interpreted in terms of K$^-$ multi-nucleon absorption processes; the possible contribution of a K$^-$pp bound state is demonstrated to overlap with the two nucleon capture process, its absolute yield thus resulting indistinguishable.

\end{abstract}

\begin{keyword}
strong interaction \sep strangeness nuclear physics \sep antikaon-nucleon interaction

\end{keyword}

\end{frontmatter}


\section{Introduction}
\label{intro}
The experimental investigation of the low-energy \kbarN interaction is crucial for the understanding of the non-perturbative QCD in the strangeness sector. The anti-kaons dynamics in-medium appears to be strongly model dependent, thus needing to be constrained by experiments. In particular, the depth of the \kbar-nucleon/nucleus optical potential has strong impact on fundamental issues, such as the partial chiral symmetry restoration with the increasing baryon density and the behaviour of kaons in cold dense nuclear matter \cite{Fuchs,Weise2015,NSR}.

Chiral perturbation theory, which is well known in the pion-nucleon case, cannot be applied to the low-energy strangeness QCD sector due to the existence of two resonances just below the \kbarN threshold (i.e. the isospin I=0 \Lambdan(1405) and the I=1 \Sigman(1385)). Phenomenological potential models predict a strongly attractive \kbarN interaction in the sub-threshold region, describing the \Lambdan(1405) as a $\bar{\mathrm{K}}$N bound state, which implies the existence of deeply bound kaonic nuclear states \cite{AY2002,IS2007,SGM2007,WG2009,Revai,MAY2013}. On the other hand, in the context of the chiral unitary SU(3) dynamics, the \Lambdan(1405) emerges as a superposition of two states (whose relative contributions are not yet known), leading to a weaker $\bar{\mathrm{K}}$N interaction and to less tightly bound kaonic nuclear states \cite{DHW2009,BGL2012,IKS2010,Bicudo,BO2013}.

From the experimental point of view the \kbar- nucleon/nucleus interaction has been studied by means of scattering experiments \cite{cs1,cs2,cs3,cs4,cs5,cs6,cs7,cs8,s0pi01,s0pi02}, kaonic atoms \cite{SIDD1,SIDD2,SIDD3,SIDD4}, measurements of kaon-nuclei absorption \cite{Katz,BC1,BC2,BC3} and search for bound kaonic states, confirming the \kbarN interaction to be attractive. 

The experimental search for kaonic bound states is controversial \cite{FINUDA2005,OBELIX,E5492008,DISTO,LEPS,HADES,E27,S0p,E15}. One of the main uncertainties when producing the bound states in K$^-$ induced reactions is represented by the lack of data concerning the K$^-$ multi-nucleon absorption processes which overlap with the bound states over a broad region of the phase spaces \cite{E5492008,OsetFINUDA,OsetLd,S0p}.

On the theoretical side, the role of the K$^-$ multi-nucleon absorption has been recently demonstrated to be fundamental in the determination of the K$^-$-nucleus optical potential. 
A phenomenological K$^-$ multi-nucleon absorption term, constrained by global absorption bubble chamber data, was added to the K$^-$ single-nucleon potential, in order to achieve good fits to K$^-$ atoms data along the periodic table \cite{FriedmanGal,Hrtankova}. An improvement in the results of the theoretical model can be obtained by characterising the absorption processes disentangling the two, three and four nucleon absorptions (2NA, 3NA and 4NA), which is one of the main results of this work.

On the experimental side, a comprehensive understanding of the K$^-$ multi-nucleon capture processes is far from being achieved: bubble chamber experiments provided only global K$^-$ multi-nucleon branching ratios (BR) \cite{Katz,BC1,BC2,BC3}; few BRs for the K$^-$ 
2NA and 3NA have been extracted within modern spectroscopic experiments \cite{S0p,FINUDA2015} and only 3NA cross sections for K$^-$ momenta of 1 GeV/c are reported in Ref. \cite{E15}. The small ($\sim$1\%) 2NA BR on deuteron compared to heavier nuclei led Wilkinson \cite{Wilkinson} to suggest a strong nuclear correlation among alpha particle-type objects at nuclear surfaces. The multi-nucleon captures involve high momentum transfer and are then ideal to test the NN  (or 3N, 4N) functions at short distances. This idea should be tested again using new data and a better understanding of nuclear theories.

Furthermore, the in-medium \kbar properties are also investigated in heavy-ion and proton-nuclei collisions. The modification of the in-medium K$^-$ mass at various densities is extrapolated from the measured K$^-$ production yield. The data are interpreted by means of transport models and collision calculations for which precise measurements of the low-energy K$^-$ multi-nucleon cross sections are mandatory (see Ref. \cite{Metag} for a complete review on this item).

This study aims to perform the first comprehensive measurement of the K$^-$ 2NA, 3NA and 4NA BRs and cross sections for low-momenta kaons in \Lambdan p and \Sigman$^0$p final states. The contribution of the possible K$^-$pp bound state is also critically investigated. 
The K$^-$ multi-nucleon absorptions are investigated by exploiting the hadronic interactions of negative kaons absorbed in $^{12}$C nuclei, the main component of the Drift Chamber (DC) inner wall of the KLOE detector \cite{KLOEcite} at the DA$\mathrm{\Phi}$NE collider \cite{DAFNEcite}. The correlated production of a \Lambdan ~ and a proton in the final state of the K$^- {}^{12}$C capture is studied.

\section{Experimental setup}
\label{sec:exp}
DA\Phin NE \cite{DAFNEcite} (Double Annular \Phin -factory for Nice Experiments) is a double ring $\mathrm{e^+ \, e^-}$ collider, designed to work at the centre of mass energy of the $\mathrm{\phi}$ particle.
The $\phi$ mesons decaying nearly at-rest produce charged kaons with BR($\mathrm{K^+ \, K^-}$) = $(48.9 \pm 0.5) \%$ at low momentum ($\sim 127$ $\mathrm{MeV/c}$), ideal either to study reactions of stopped kaons, or to explore the products of the low-energy nuclear absorptions of K$^-$. The back-to-back topology of the emitted K$^-$K$^+$ pairs allows to tag the K$^-$ through the detection of a K$^+$ in the opposite direction, if the K$^-$ track is not identified. 

The KLOE detector \cite{KLOEcite} is centred around the interaction point of DA\Phin NE. KLOE is characterised by a $\sim 4\pi$ geometry and an acceptance of $\sim98\%$; it consists of a large cylindrical Drift Chamber (DC) \cite{KLOEdc} and a fine sampling lead-scintillating fibres calorimeter \cite{KLOEemc}, all immersed in an axial magnetic field of 0.52 $\mathrm{T}$ provided by a superconducting solenoid.
The DC has an inner radius of 0.25 $\mathrm{m}$, an outer radius of 2 $\mathrm{m}$ and a length of 3.3 $\mathrm{m}$. The DC inner wall composition is 750 $\mathrm{\mu m}$ of Carbon fibre and 150 $\mathrm{\mu m}$ of Aluminium foil.
The KLOE DC is filled with a mixture of Helium and Isobutane (90$\%$ in volume $^4\textrm{He}$ and 10$\%$ C$_4$H$_{10}$). 
The chamber is characterised by excellent position and momentum resolutions: tracks are reconstructed with a resolution in the transverse $R-\phi$ plane
$\sigma_{R\phi}\sim200\,\mathrm{\mu m}$ and a resolution along the z-axis $\sigma_z\sim2\,\mathrm{mm}$, while the transverse momentum resolution for low-momentum tracks ($(50<p<300) \mathrm{MeV/c}$)
is $\frac{\sigma_{p_T}}{p_T}\sim0.4\%$.

The KLOE calorimeter is composed of a cylindrical barrel and two endcaps, providing a solid angle coverage of 98\%.
The volume ratio (lead/fibres/glue = 42:48:10) is optimised for
a high light yield and a high efficiency for photons in the range
(20-300) MeV/c. The photon detection efficiency is 99$\%$ for energies larger than 80 MeV and it falls to 80$\%$ at 20 MeV due to the cutoff introduced by the readout  
threshold. The position of the clusters along the fibres can be obtained with a resolution $\sigma_{\parallel} \sim 1.4\, \mathrm{cm}/\sqrt{E(\mathrm{GeV})}$. The resolution in the orthogonal direction is  $\sigma_{\perp} \sim 1.3\, \mathrm{cm}$. The energy and time resolutions for photon clusters are given by $\frac{\sigma_E}{E_\gamma}= \frac{0.057}{\sqrt{E_\gamma (\mathrm{GeV})}}$ and 
$\sigma_t= \frac{57 \, \mathrm{ps}}{\sqrt{E_\gamma (\mathrm{GeV})}} \oplus 100 \,\, \mathrm{ps}$.

An integrated luminosity of 1.74 fb$^{-1}$ collected by the KLOE collaboration in the 2004/2005 data campaign was analysed. The inner wall of the KLOE DC is used as target. Dedicated GEANT Monte Carlo (MC) simulations of the KLOE apparatus (GEANFI \cite{GEANFI}) were performed to estimate the percentages of K$^-$ absorptions in the materials of the DC inner wall (the K$^-$ absorption physics is treated by the GEISHA package). Out of the total fraction of captured kaons, about 81$\%$ is absorbed in the Carbon fibre component and the residual 19$\%$ in the Aluminium foil. Motivated by the larger amount of K$^-$ captures in the Carbon fibre component and due to the similarity of the kinematic distributions for K$^-$ absorptions in $^{12}$C and $^{27}$Al, not distinguishable within the experimental resolution, a pure Carbon target was considered in the analysis.

The analysed data include both K$^-$ captures at-rest and in-flight. In the first case the negatively charged kaon is slowed down in the materials of the detector and consequently absorbed in a highly excited atomic orbit, from which it cascades down till it is absorbed by the nucleus through the strong interaction. In the latter case the kaon penetrates the electronic cloud and interacts with the nucleus with an average momentum of about $100$ MeV/c.

\section{\Lambdan p events selection} 
\label{sec:evselection}
The starting point of the analysis of K$^-$ absorption processes is the identification of a \Lambdan$ (1116)$ hyperon through its decay into a proton and a negatively charged pion (BR = (63.9 $\pm$ 0.5) \% \cite{PDG}).

The proton and pion tracks are identified using the dE/dx information measured in the DC wires, together with the energy information provided by the calorimeter.
For each pion and proton candidate a minimum track length of 30 cm is required. Additionally, the proton candidate must have a momentum higher than $p >\,170$ MeV/c in order to minimise the pion contamination in the proton sample and to improve the efficiency of the particle identification.
A common vertex is searched for all the p$\pi^-$ pairs.
The reconstructed invariant mass $m_{p\pi^-}$ shows a mean value of (1115.753 $\pm$ 0.002) MeV/c$^2$ and a resolution of $\sigma$ = 0.5 MeV/c$^2$.
For the next analysis steps, the following cut is applied on the p$\pi^-$ invariant mass: $(1112 < m_\mathrm{p \pi^-} < 1118)$ MeV/c$^2$.

After the \Lambdan ~decay vertex selection, the K$^-$ hadronic interaction vertex is reconstructed. To this end events with an additional proton track are selected by applying the same cuts as for the protons associated to the \Lambdan ~decay. We will refer to the additional proton as ``primary proton'', to stress that this proton and the \Lambdan~ are the primary products of the K$^-$ hadronic absorption. The vertex position is determined by backward extrapolation of the \Lambdan ~path and the primary proton track with a resolution of 0.12 cm. At the point of closest approach, a distance smaller than 5 cm is required. Particles momenta are corrected for energy loss.

The distribution of the \Lambdan p (hadronic) vertices along the radial direction ($\rho_\mathrm{\Lambda p}$), orthogonal with respect to the DA\Phin NE beam pipe, is used to select the K$^-$ absorptions occurring in the DC inner wall. 
The geometrical cut $\rho_\mathrm{\Lambda p} = (25.0 \pm 1.2 )\ \mathrm{cm}$ was chosen to minimise the contamination from interactions with the gas of the DC \cite{Tesikri}. 

The goal of this analysis is to extract the branching ratios and the low-energy cross sections for the K$^-$ absorptions on two, three and four nucleons (2NA, 3NA and 4NA, respectively), which are non-pionic processes: 
\begin{equation}
\mathrm{K^- (NN) \rightarrow Y N \qquad (2NA) \ ,} \nonumber
\label{multi-nucleon1}
\end{equation}
\begin{equation}
\mathrm{K^- (NNN) \rightarrow Y ``NN" \qquad (3NA) \ ,}
\label{multi-nucleon2}
\end{equation}
\begin{equation}
\mathrm{K^- (NNNN) \rightarrow Y ``NNN" \qquad (4NA) \ ,} \nonumber
\label{multi-nucleon3}
\end{equation}
where (NN), (NNN) and (NNNN) are bound nucleons in the target nucleus, Y is the hyperon, ``NN" and ``NNN" represent the final state nucleons which can be bound or unbound. The main background source is represented by the pionic K$^-$ single nucleon absorption processes:
\begin{equation}
\mathrm{K^- (N) \rightarrow Y \pi \qquad (1NA) \ ,}
\end{equation}
that always lead to the production of a hyperon and a pion in the final state and whose global BR in $^{12}$C is found to be about 80 \% in emulsion experiments \cite{BC3}. 

\begin{figure}[!h]
\centering
  \includegraphics[width=0.7\textwidth]{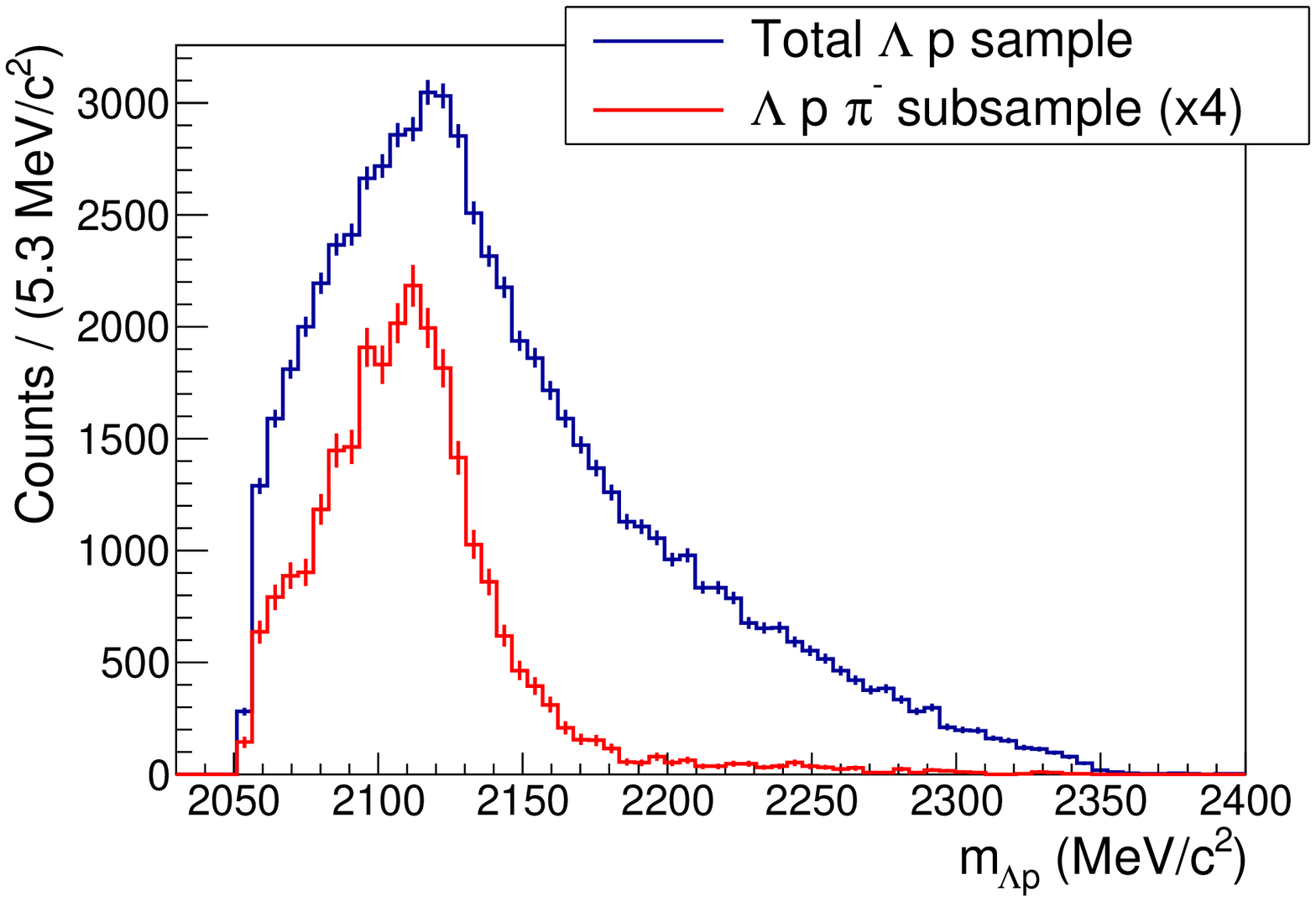}
  \includegraphics[width=0.7\textwidth]{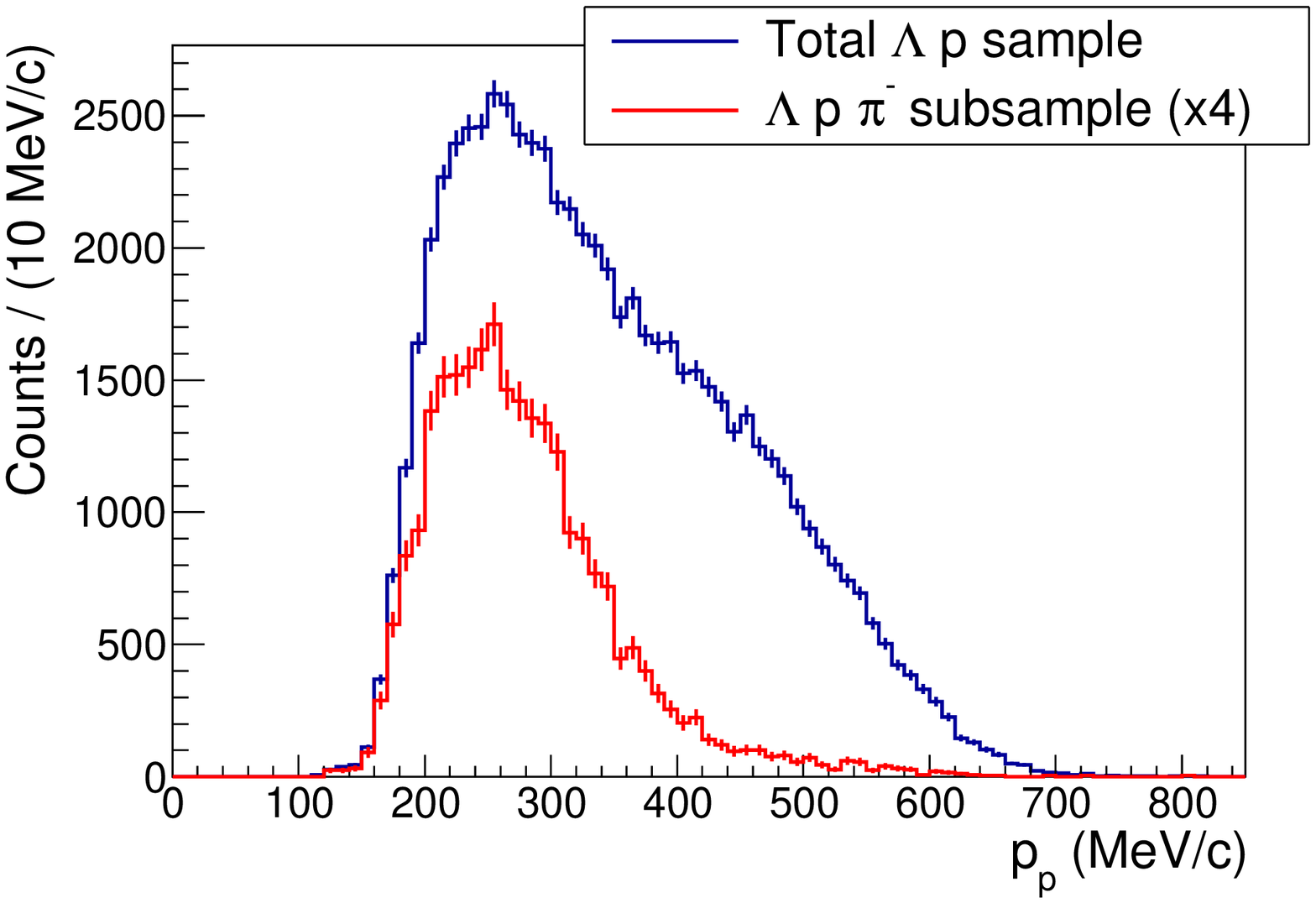}
\caption{\Lambdan p invariant mass (top) and proton momentum (bottom) distributions are shown for the whole \Lambdan p data sample (blue) together with the \Lambdan p$\pi^-$ subsample events (red). The red distribution is multiplied by a factor 4 for a more clear comparison.}
\label{fig:1NAb}       
\end{figure}

In the sample of the reconstructed \Lambdan p events an additional $\pi^-$ or $\pi^+$ correlated with the \Lambdan p pair at the hadronic vertex, representing the signature of the K$^-$ absorption on a single nucleon, is found in 16.3\% and 6.1\% of the total events, respectively. In these cases the proton which is found to be correlated with the \Lambdan ~production does not originate in the primary K$^-$ hadronic absorption; it results from final state interaction (FSI) processes with fragmentation of the residual nucleus. Such events contribute to the lower region of the \Lambdan p invariant mass distribution and are characterised by low-momenta protons, as can be seen in Fig.~\ref{fig:1NAb}.
Since these protons originate from the residual nucleus fragmentation, most of them are Fermi sea nucleons (the Fermi momentum distribution in $^{12}$C is centred around 220 MeV/c), where the momentum distribution is partly distorted by the FSI processes.

\begin{figure}
\centering
  \includegraphics[width=0.7\textwidth]{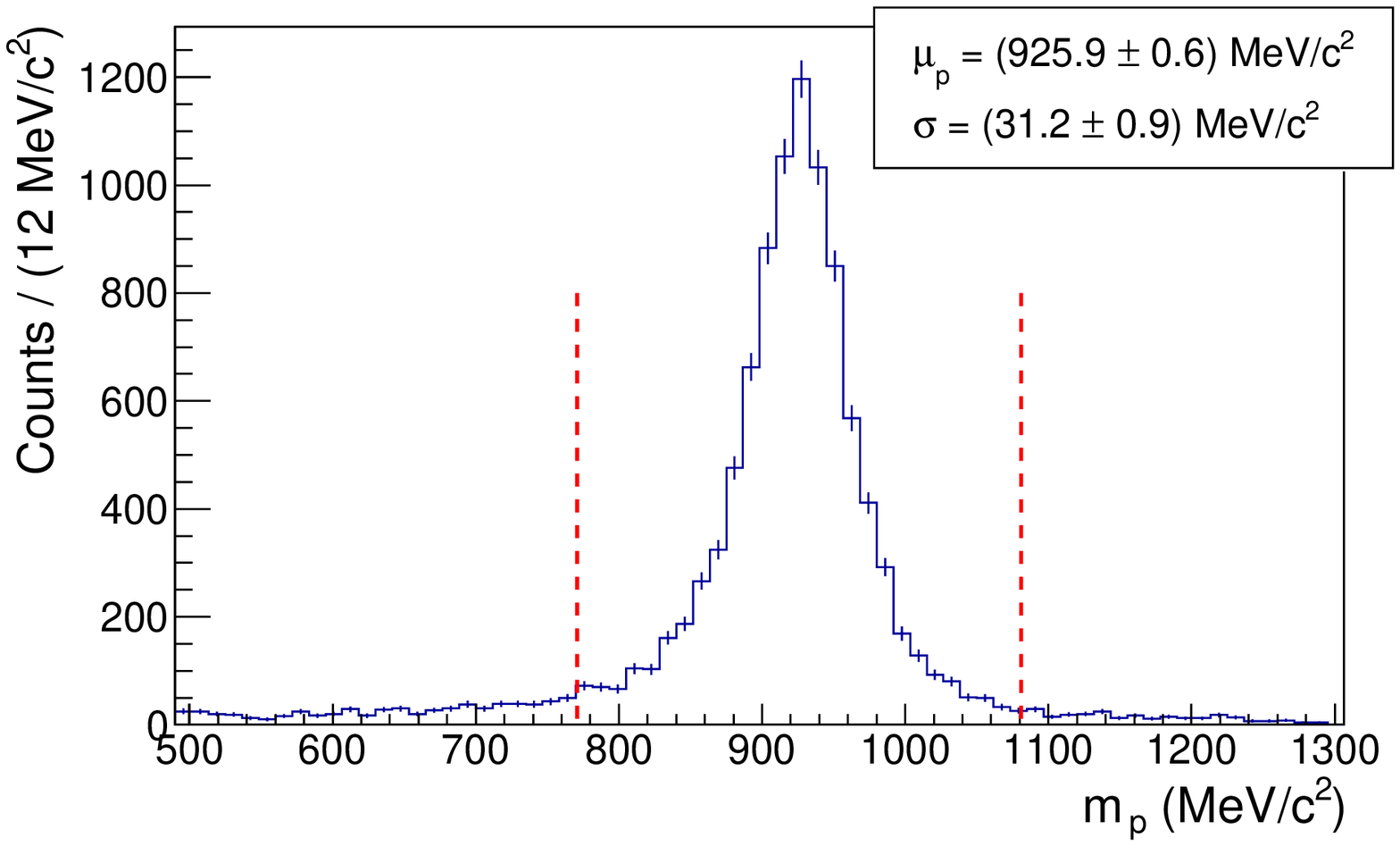}
  \includegraphics[width=0.7\textwidth]{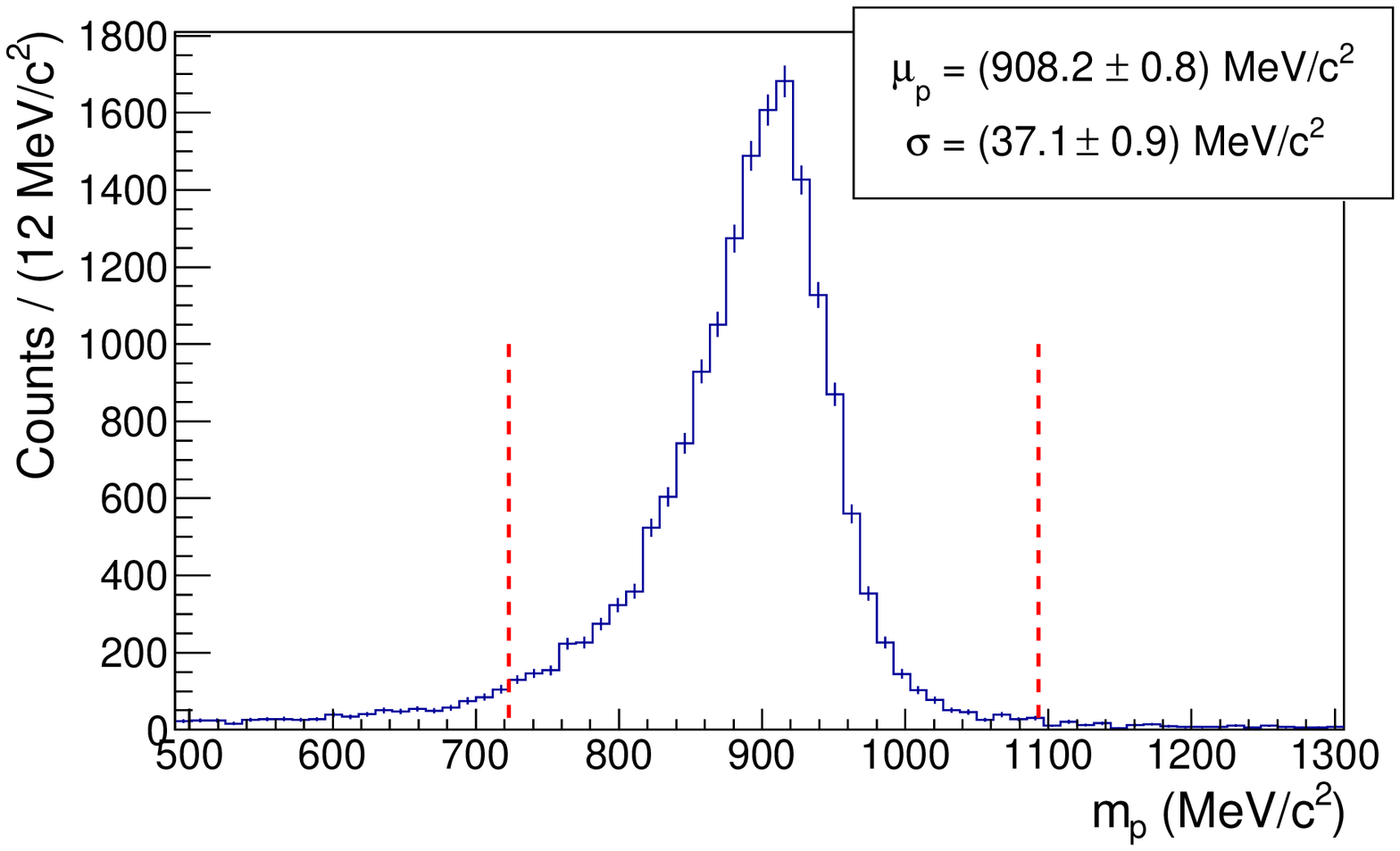}
\caption{Mass of the proton from the \Lambdan ~decay (top) and the primary proton mass (bottom) measured by TOF. The red dashed lines represent the applied cut.}
\label{fig:pmass}       
\end{figure}

As it will be shown at the end of this Section, the contribution of the K$^-$ single nucleon absorptions to the selected \Lambdan p sample is strongly reduced when the measurement of the mass of both selected protons, by using the time of flight (TOF) information, is required, allowing to isolate an almost pure sample of multi-nucleon K$^-$ captures. 
The mass distributions of the two protons are shown in Fig.~\ref{fig:pmass}. 
A Gaussian fit of the peaks (neglecting the energy loss tails) is performed and the cut $m_\mathrm{p} = \mathrm{(\mu_p \pm 5 \sigma)}$ is applied for the selection of the final protons samples ($\mu_\mathrm{p}$ and $\sigma$ represent the mean value and the standard deviation obtained from the fit and are reported in Fig. \ref{fig:pmass}). The mean values of the two mass distributions are smaller than the nominal mass of the proton as a consequence of energy loss effect. Such effect is more pronounced for the primary protons since they are produced in the DC inner wall, while most of the \Lambdan s decay within the DC volume. 

\begin{figure}[!h]
\centering
  \includegraphics[width=0.7\textwidth]{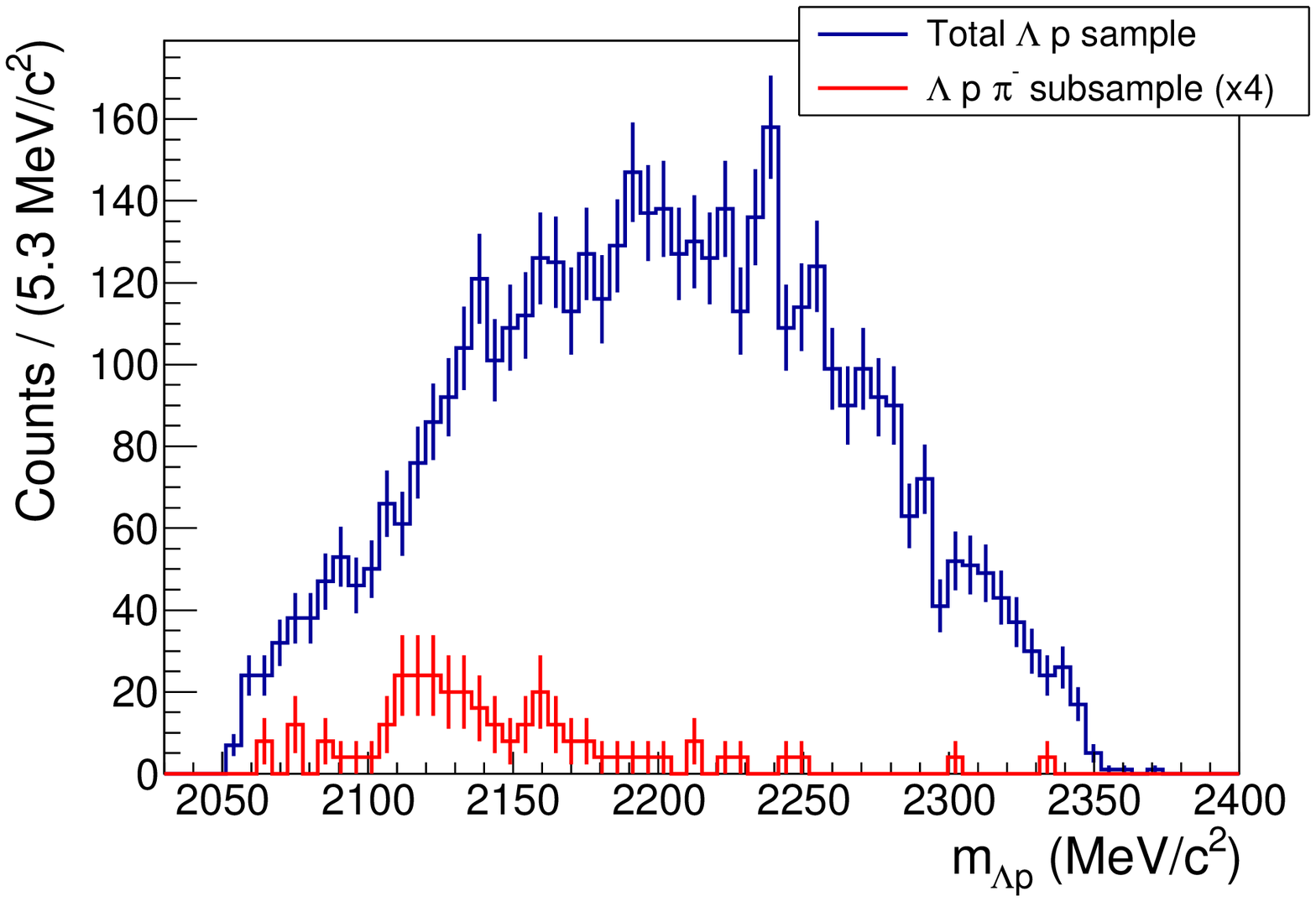}
  \includegraphics[width=0.7\textwidth]{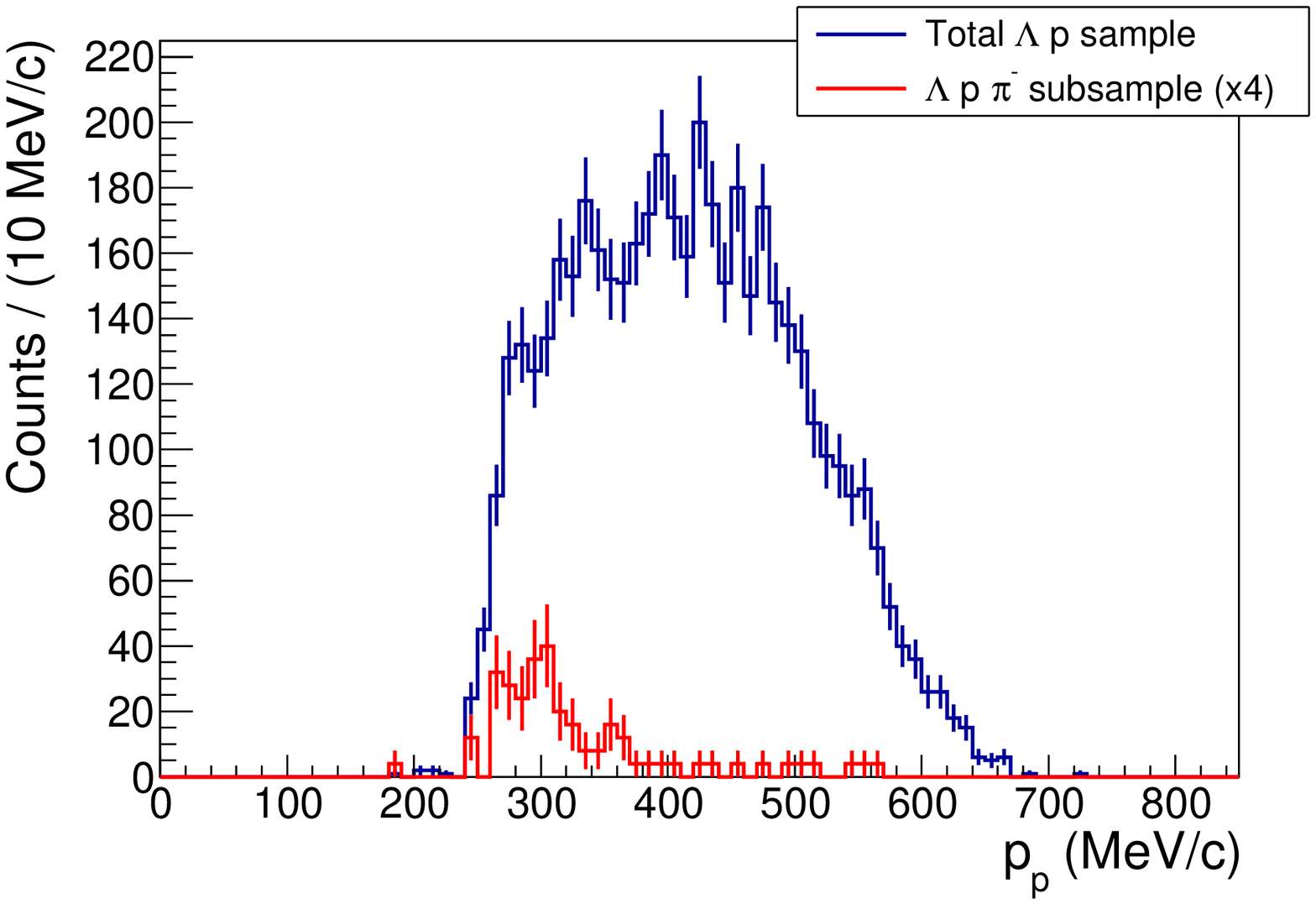}
\caption{\Lambdan p invariant mass (top) and proton momentum (bottom) distributions after the selection of the protons from the mass by TOF are shown for the whole \Lambdan p data sample (blue) together with the \Lambdan p$\pi^-$ subsample events (red). The red distribution is multiplied by a factor 4 for
a more clear comparison.}
\label{fig:1NAa}   
\end{figure}

In order to measure the mass of the protons by TOF both protons must have an associated cluster in the calorimeter. Since the energy threshold of the KLOE calorimeter corresponds to a proton momentum of about 240~MeV/c, larger than the Fermi momentum in $^{12}$C, the contamination of pionic processes is reduced to 2 \% in the final \Lambdan p sample.  
The contribution of single nucleon absorption processes will not be considered in the fit, but will be taken into account into the systematic errors.
The comparison between the \Lambdan p invariant mass and proton momentum distributions of the \Lambdan p$\pi^-$ and the total \Lambdan p sample, after the selection of the protons by using the TOF is shown in Fig. \ref{fig:1NAa}.

3 \% of the events belonging to the lower region of the proton momentum distribution ($p_\mathrm{p} <$ 270 MeV/c) were removed in order to avoid biases due to the calorimeter threshold for low energy protons.

In order to extract the BRs and the cross sections of the K$^-$ multi-nucleon absorption processes, the final event selection has been performed also requiring the tagging of the K$^-$ through the detection of a K$^+$ in the opposite hemisphere of the KLOE detector. 

The final number of reconstructed \Lambdan p events is 4543, which was then analysed.

\section{Monte Carlo simulations}
\label{sec:MC}
The processes which contribute to \Lambdan p production in the final state of K$^-$ absorptions in $^{12}$C are simulated with the aim to fit the measured distributions in order to extract the corresponding BRs and cross sections.
Beside the direct \Lambdan ~production in the final state, also the primary \Sigman$^0$ production followed by the \Sigman$^0 \rightarrow$ \Lambdan $\gamma$ electromagnetic decay is to be accounted for.

The 2NA, 3NA and 4NA processes reported in Eqs.~(\ref{multi-nucleon1}-\ref{multi-nucleon3}) are simulated in both the \Lambdan p and \Sigman$^0$p channels.
For the 2NA two contributions can be distinguished: 1) the quasi-free (QF) production of the \Lambdan(\Sigman$^0$)p pairs, without FSIs with the residual nucleus; 2) the primary \Lambdan(\Sigman$^0$)N production followed by elastic FSIs (in this case only single collisions of the hyperon or the nucleon with the residual nucleus are accounted for). Furthermore the inelastic $\mathrm{\Sigma N \rightarrow \Lambda N'}$ scattering reactions with the Fermi sea residual nucleons remaining after the primary interaction are also considered leading to the conversion of the \Sigman s into \Lambdan~ hyperons.

In case of 2NA-QF processes 
\begin{eqnarray}
\mathrm{K^- + {}^{12}C \rightarrow K^- + (pp) + ^{10}Be \rightarrow \Lambda(\Sigma^0)  + p + ^{10}Be} \ ,
\nonumber
\end{eqnarray}
the possibility for the residual ${}^{10}$Be nucleus to be left in an excited state configuration is analysed. This leads to the ${}^{10}$Be fragmentation essentially reflected in a slight lowering of the \Lambdan p invariant mass. 

The relative amplitudes of all the mentioned processes are free parameters of the global fit.

The kinematics of each process, defined by the hyperon Y (\Lambdan ~or \Sigman$^0$) and the proton vector momenta ($\mathrm{\textbf{\textit{p}}_{Y}, \textbf{\textit{p}}_p}$) are generated by sampling the total hyperon-proton momentum ($\mathrm{\textbf{\textit{p}}_{Yp}}$) distribution $P(\mathrm{\textbf{\textit{p}}_{Yp}})$, applying energy and momentum conservation. 

The ($\mathrm{\textbf{\textit{p}}_{Y}, \textbf{\textit{p}}_p}$) pair represents the input for the GEANFI digitisation followed by the event reconstruction. The momentum distributions are obtained following the phenomenological K$^-$ absorption model described in Refs. \cite{io,PisWycCur}:
\begin{equation}
P(\mathrm{\textit{p}_{Yp}})\mathrm{d\textit{p}_{Yp}} = \mathrm{|T(\textit{m}_{Yp})|^2 \cdot |F(\textit{p}_{Yp})|^2 \cdot d\rho \ ,}
\end{equation}
the (unknown) energy dependence of the K$^-$ multi-nucleon absorption transition amplitudes ($\mathrm{T(\textit{m}_{Yp})}$) is neglected, $\mathrm{T(\textit{m}_{Yp})}$ is then assumed uniform, the value has to be determined experimentally from the fit. $\mathrm{F(\textit{p}_{Yp})}$ is the form-factor containing ``all'' the nuclear physics, defined as in Ref. \cite{io} replacing the single proton wave function with the two, three and four nucleons wave function for the 2NA, 3NA and 4NA processes, respectively. $\mathrm{d\rho}$ is the phase space element and includes also the recoil of the residual nucleus. Gaussian wave functions are used for nucleons bound in the $^{12}$C nucleus to calculate $\mathrm{F(\textit{p}_{Yp})}$, according to Refs. \cite{Henley,nucldensity}. The wave function of the K$^-$ depends on the absorption process. 
For the at-rest K$^-$ capture the kaons are assumed to be absorbed from the 2p atomic level in $^{12}$C according to \cite{OsetFINUDA}; in this work the dependence of the \Lambdan p invariant mass shape from the assumed atomic angular momentum state (3d or 2p) is found to be negligible. For the in-flight K$^-$ capture the wave function adopted in Ref. \cite{PisWycCur} is used. 

In the simulations the same amount of K$^-$ captures at-rest and in-flight is considered, based on Ref. \cite{lpi}; in this work the direct \Lambdan$\pi^-$ production following K$^-$ absorptions on neutrons is investigated considering the separate contributions of K$^-$ at-rest and in-flight captures, which are found to equally contribute (within the experimental errors) to the final measured sample. In the present work, the error introduced by this assumption was included as a contribution to the systematic errors.
The modulus of the K$^-$ momentum, used to simulate the in-flight captures, is sampled according to the measured momentum distribution at the last point of the K$^-$ track when the hadronic absorption occurs and a $\mathrm{\Lambda}$ is detected, according to the selection described in Section 2.

\section{Fit of the data}
\label{sec:fit}
The function used in the fit to model the measured kinematic distributions is taken as a linear combination of the MC components for the contributing processes previously described in Section \ref{sec:MC}, and assumes the following expression:
\begin{equation}
\mathcal{F}^q (q_n) = \sum\limits_{i=1}^{N_{par}} \alpha_i \cdot h_i^{ q} (q_n) \ ,
\end{equation}
where $\alpha_i$s are the free parameters of the fit, $h_i^{ q}$ represent the distribution of the kinematic variable $q$, corresponding to the $i$-th process, normalised to the data entries, $N_{par}$ is the total number of parameters and $q_n$ is the $n$-th bin for the variable $q$.

\begin{figure*}[!h]
\centering
\includegraphics[width=0.49\textwidth]{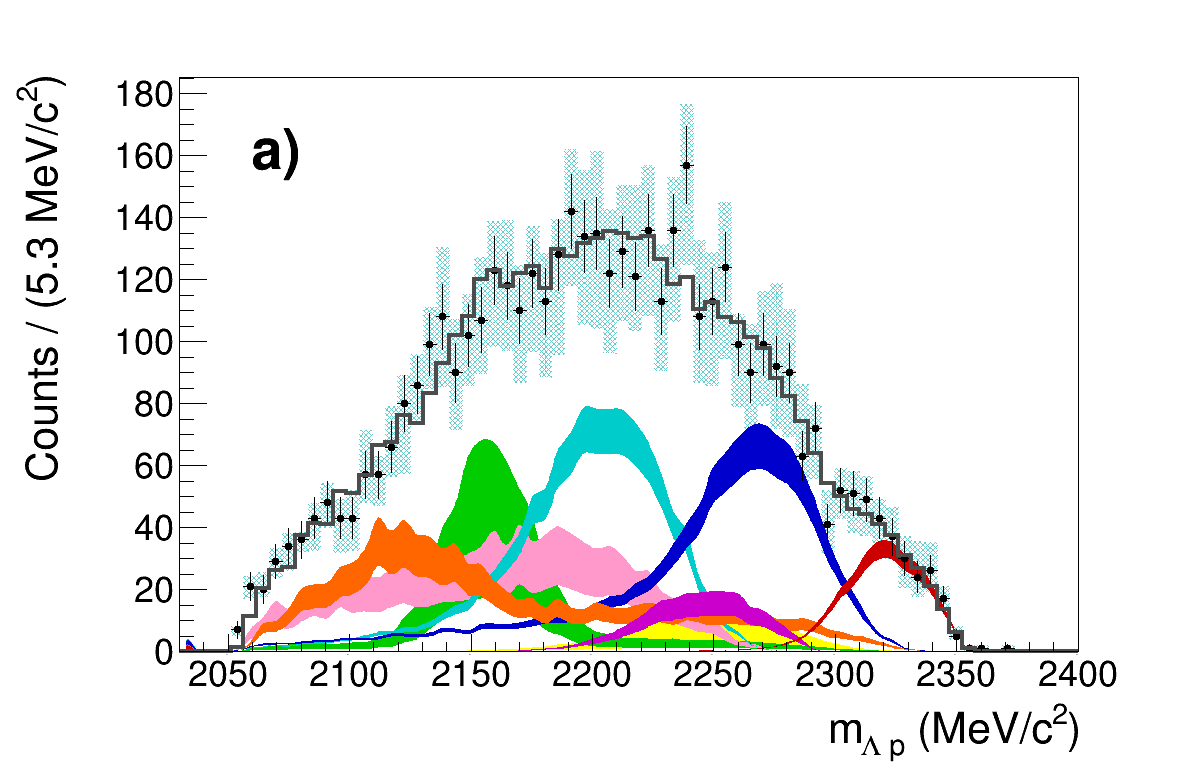}
\includegraphics[width=0.49\textwidth]{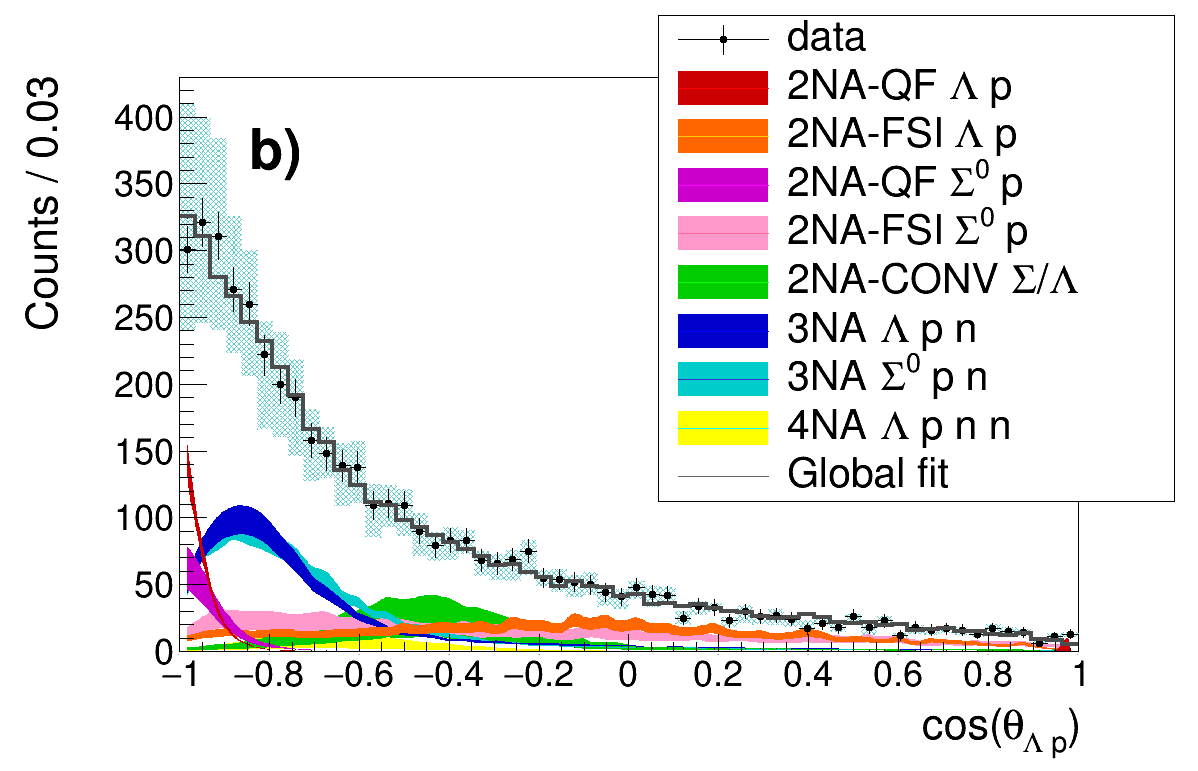}
\includegraphics[width=0.49\textwidth]{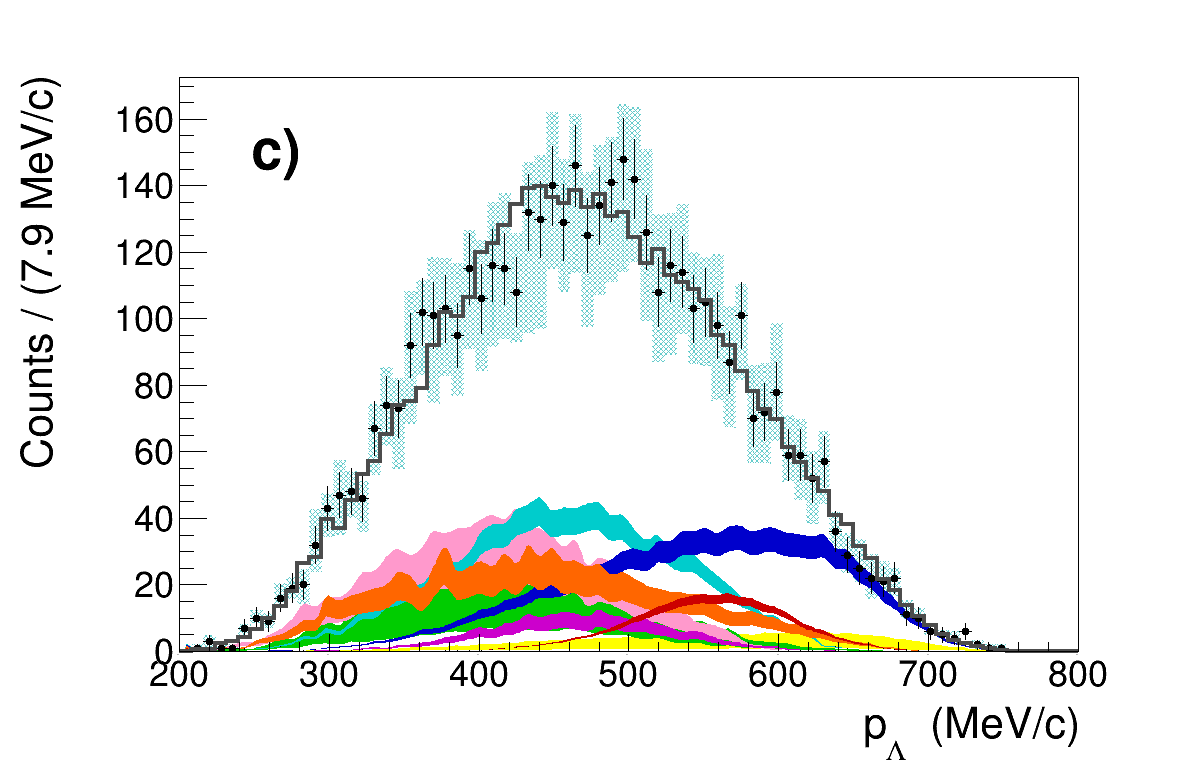}
\includegraphics[width=0.49\textwidth]{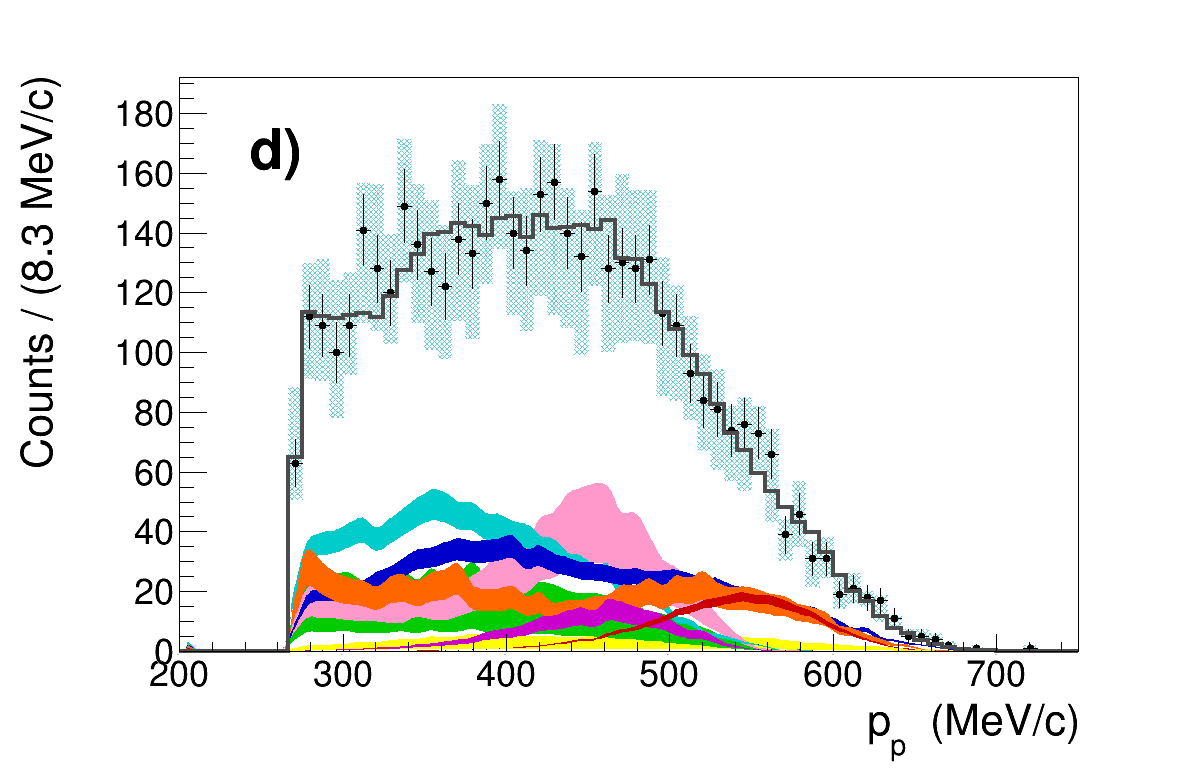}
\caption{\Lambdan p invariant mass (panel a), \Lambdan p angular correlation (panel b), \Lambdan ~momentum (panel c) and proton momentum (panel d) distributions are reported for the K$^-$ absorption on $^{12}$C listed in the legend. Black points represent the data, black error bars correspond to the statistical errors, cyan error bars correspond to the systematic errors. The gray line distributions represent the global fitting functions, the coloured distributions represent the different contributing processes according to the colour code reported in the legend and the widths correspond to the statistical error.}
\label{fit}
\end{figure*}

A simultaneous $\chi^2$ fit of the \Lambdan p invariant mass ($m_\mathrm{\Lambda p}$), \Lambdan p angular correlation ($\cos\theta_\mathrm{\Lambda p}$), \Lambdan ~momentum ($p_\mathrm{\Lambda}$) and proton momentum ($p_\mathrm{p}$) is performed using the SIMPLEX, MIGRAD and MINOS routines of ROOT \cite{MINUIT1,MINUIT2,MINUIT3} for the minimisation procedure. 
The result of the fit is shown in Fig. \ref{fit}. 
Black points represent the data, black error bars correspond to the statistical errors and cyan error bars correspond to the systematic errors obtained according to the procedure that will be described in Section \ref{sec:syst}. The gray line distributions represent the global fitting functions, the coloured distributions represent the various contributing processes and their widths correspond to the statistical error. The reduced chi-squared is $\chi^2/dof = 194 / 206 = 0.94$.

For the evaluation of BRs and cross sections of the various processes the corresponding detection efficiencies, obtained on the basis of the MC simulations, are used. The larger efficiencies are obtained for the 2NA-QF and for the 3NA. The efficiency of the 2NA-QF is found to be (2.49 $\pm$ 0.02) \% in the \Lambdan p channel and (1.24 $\pm$ 0.26) \% in the \Sigman$^0$p channel, while those of the 3NA are (1.77 $\pm$ 0.01) \% and (0.65 $\pm$ 0.01) \%, respectively.  
The detection efficiency is energy dependent and increases with the track momentum, thus it is higher for the K$^-$ 2NA-QF in the \Lambdan p channel since the protons are emitted with a larger mean momentum with respect to the other processes, as shown in Fig. \ref{fit}. 
The obtained efficiencies are lower than 3\% due to the protons selection using the mass by TOF. The measurement of the TOF, indeed, requires the energy cluster association in the calorimeter for the two proton tracks, reducing the amount of selected events.

The BR for the process $i$ is defined as
\vspace{1mm}
\begin{equation}
\mathrm{BR_i = \frac{N_i^{at-rest}}{N_{K^- stop}} \ ,}
\end{equation}
\vspace{1mm}
where $\mathrm{N_i^{at-rest}}$ represents the absolute number of events for the $i$ absorption process when the K$^-$ is captured at-rest, obtained from the corresponding contribution to the fit corrected by the detection efficiency; $\mathrm{{N_{K^- stop}}}$ is the number of stopped K$^-$ in the DC inner wall for the analysed luminosity which have been tagged through the detection of a K$^+$, emitted in the opposite hemisphere of the KLOE detector.  

The cross sections are evaluated as 
\begin{equation}
\mathrm{\sigma_i = \frac{N_i^{in-flight}}{N_{K^-}^{projectiles} \cdot n \cdot L} \ ,}
\end{equation}
where $\mathrm{{N_i^{in-flight}}}$ represents the absolute number of events for the $i$ absorption process when the K$^-$ is captured in-flight, obtained from the corresponding contribution to the fit corrected by the detection efficiency; $\mathrm{N_{K^-}^{projectiles}}$ is the number of K$^-$ impinging on the DC inner wall for the analysed luminosity which have been tagged;  
n is the density of the interaction centres of the target material and L is the thickness of the Carbon wall.
The obtained BRs and cross sections are summarised in Table \ref{BRandCS}. The K$^-$ momentum at which the cross section is measured is evaluated in the centre of mass reference frame of the absorbing nucleons, thus it differs for the 2NA and 3NA processes.
\begin{table*}[h]
\caption{Branching ratios (for the K$^-$ absorbed at-rest) and cross sections (for the K$^-$ absorbed in-flight) of the K$^-$ multi-nucleon absorption processes. The K$^-$ momentum is evaluated in the centre of mass reference frame of the absorbing nucleons, thus it differs for the 2NA and 3NA processes. The statistical and systematic errors are also given.}
\label{BRandCS}
\resizebox{\textwidth}{!}{
\begin{tabular}{ll|ll}
\hline
Process & \multicolumn{1}{l|}{Branching Ratio (\%)} & \multicolumn{1}{l}{$\sigma$ (mb)} & \multicolumn{1}{l}{@ ~~~~~~ $p_K$ (MeV/c)}\\
\hline
2NA-QF \Lambdan p & 0.25 $\pm$ 0.02 (stat.) ${}^{+0.01}_{-0.02}$(syst.) & 2.8 $\pm$ 0.3 (stat.) ${}^{+0.1}_{-0.2}$ (syst.) & @ ~~~~~~ 128 $\pm$ 29 \\[0.5ex]
2NA-FSI \Lambdan p &  6.2 $\pm$ 1.4(stat.) ${}^{+0.5}_{-0.6}$(syst.) & 69 $\pm$ 15 (stat.) $\pm$ 6 (syst.) & @ ~~~~~~ 128 $\pm$ 29 \\[0.5ex]
2NA-QF \Sigman$^0$p &  0.35 $\pm$ 0.09(stat.) ${}^{+0.13}_{-0.06}$(syst.) &  3.9 $\pm$ 1.0 (stat.)  ${}^{+1.4}_{-0.7}$ (syst.) & @ ~~~~~~ 128 $\pm$ 29 \\[0.5ex]
2NA-FSI \Sigman$^0$p &  7.2 $\pm$ 2.2(stat.) ${}^{+4.2}_{-5.4}$(syst.)   & 80 $\pm$ 25 (stat.) ${}^{+46}_{-60}$ (syst.) & @ ~~~~~~ 128 $\pm$ 29 \\[0.5ex]
2NA-CONV \Sigman /\Lambdan &  2.1 $\pm$ 1.2(stat.) ${}^{+0.9}_{-0.5}$(syst.) & -  \\[0.5ex]
3NA \Lambdan pn &   1.4 $\pm$ 0.2(stat.) ${}^{+0.1}_{-0.2}$(syst.)  & 15 $\pm$ 2 (stat.) $\pm$ 2 (syst.) & @ ~~~~~~ 117 $\pm$ 23 \\[0.5ex]
3NA \Sigman$^0$pn &  3.7 $\pm$ 0.4(stat.) ${}^{+0.2}_{-0.4}$(syst.) & 41 $\pm$ 4 (stat.) ${}^{+2}_{-5}$ (syst.) & @ ~~~~~~ 117 $\pm$ 23 \\[0.5ex]
4NA \Lambdan pnn &  0.13 $\pm$ 0.09(stat.) ${}^{+0.08}_{-0.07}$(syst.)   & - \\[0.5ex]
\hline
Global \Lambdan(\Sigman$^0$)p & 21 $\pm$ 3(stat.) ${}^{+5}_{-6}$(syst.)  &  - \\[0.5ex]
\hline
\end{tabular}}
\end{table*}
\section{Systematic errors}
\label{sec:syst}
The systematic uncertainties for the measurements are obtained by performing variations of each cut applied for the \Lambdan p event selection described in Section \ref{sec:evselection}. The systematic error to the parameter $\alpha_i$ introduced by the variation of the $j$-th cut is given by:
\begin{equation}
\sigma_{sist., i}^{j} = \alpha_i^{j} - \alpha_i \ , 
\end{equation}
where $\alpha_i^j$ is the value obtained from the new fit. The total, positive and negative, systematic errors are obtained summing in quadrature, separately, the positive and negative systematics.

In Table \ref{tabcuts}, the standard selection cuts for the kinematic variables which are found to significantly contribute to the systematic errors are summarised.  
\begin{table*}[!h]
\centering
\caption{Cuts of the optimised \Lambdan p event selection that significantly contribute to the systematic errors. See Fig. \ref{fig:pmass} for further details about the cuts applied to the mass of the protons.}
\resizebox{0.9\linewidth}{!}{
\begin{tabular}{cc}
\hline \noalign{\vskip 0.05in}
Hadronic vertex radial coordinate: & 23.8 cm $< \rho_\mathrm{\Lambda p} <$ 26.2 cm\\ [0.5ex]
\hline \noalign{\vskip 0.05in}
Primary proton mass: &  723 MeV/c$^2$ $< m_\mathrm{p}  <$ 1093 MeV/c$^2$\\[0.5ex]
\hline \noalign{\vskip 0.05in}
Proton from the $\mathrm{\Lambda}$ decay mass: &  771 MeV/c$^2$ $< m_\mathrm{p}  <$ 1081 MeV/c$^2$\\[0.5ex]
\hline \noalign{\vskip 0.05in}
Primary proton momentum: & $p_\mathrm{p} \geq 270$ MeV/c \\[0.5ex]
\hline
\end{tabular}
}
\label{tabcuts}
\end{table*}
The $\rho_\mathrm{\Lambda p}$ cut variations are performed in order to select $\pm$15\% of events. For the proton TOF mass cuts we relied on the standard deviation of the mass distribution fits ($\pm 2 \sigma$) and the systematic on the proton momentum cut is evaluated by removing it.
Two important additional contributions to the systematic errors are also considered, related to the assumptions of the event modelling, namely the ansatz on the equal in-flight/at-rest contribution, which is varied to 40/60\% and 60/40\%, and the contribution of the single nucleon absorptions, evaluated performing the fit with additional pionic contributions. 

\section{Discussion of the results}
\label{sec:discussion}
\subsection{K$^-$ multi-nucleon absorption BRs and cross sections}
The BRs and cross sections of the 2NA-QF and of the 3NA with the production of the \Lambdan ~and the \Sigman$^0$ in the final state are obtained with a better precision with respect to the other components, as it is evident from Fig.~\ref{fit} and Table \ref{BRandCS}. For these processes the relative error (combined statistical and systematic) on the measured yields is smaller than 20\% except for the 2NA-QF \Sigman$^0$p. 

The fit is not sensitive to the contribution of the 4NA process due to the overlap with the 3NA over a broad range of the phase space (compare for example the invariant mass distributions in Fig. \ref{fit}); moreover the 4NA \Lambdan p production is strongly uncorrelated, similar to FSI or conversion processes (as evident from the $\cos\theta_\mathrm{\Lambda p}$ spectra). Consequently, is not possible to pin down the 4NA with enough accuracy.
The 2NA-QF component with the direct \Lambdan p production is clearly disentangled from the other contributions, especially in the \Lambdan p invariant mass distribution. The 3NA can also be well identified and is measured with unprecedented accuracy for the \Lambdan p channel at low-energy and for the \Sigman$^0$p channel \cite{E15}. 
The global branching ratio of the K$^-$ multi-nucleon absorptions in ${}^{12}$C producing \Lambdan(\Sigman$^0$)p final states, obtained summing all the BRs of Table \ref{BRandCS}, is found to be consistent with the bubble chamber measurements \cite{FriedmanGal}.

\subsection{K$^-$p coupling to the \Lambdan(1405)}
\begin{figure}[!h]
\centering
\includegraphics[width=0.7\textwidth]{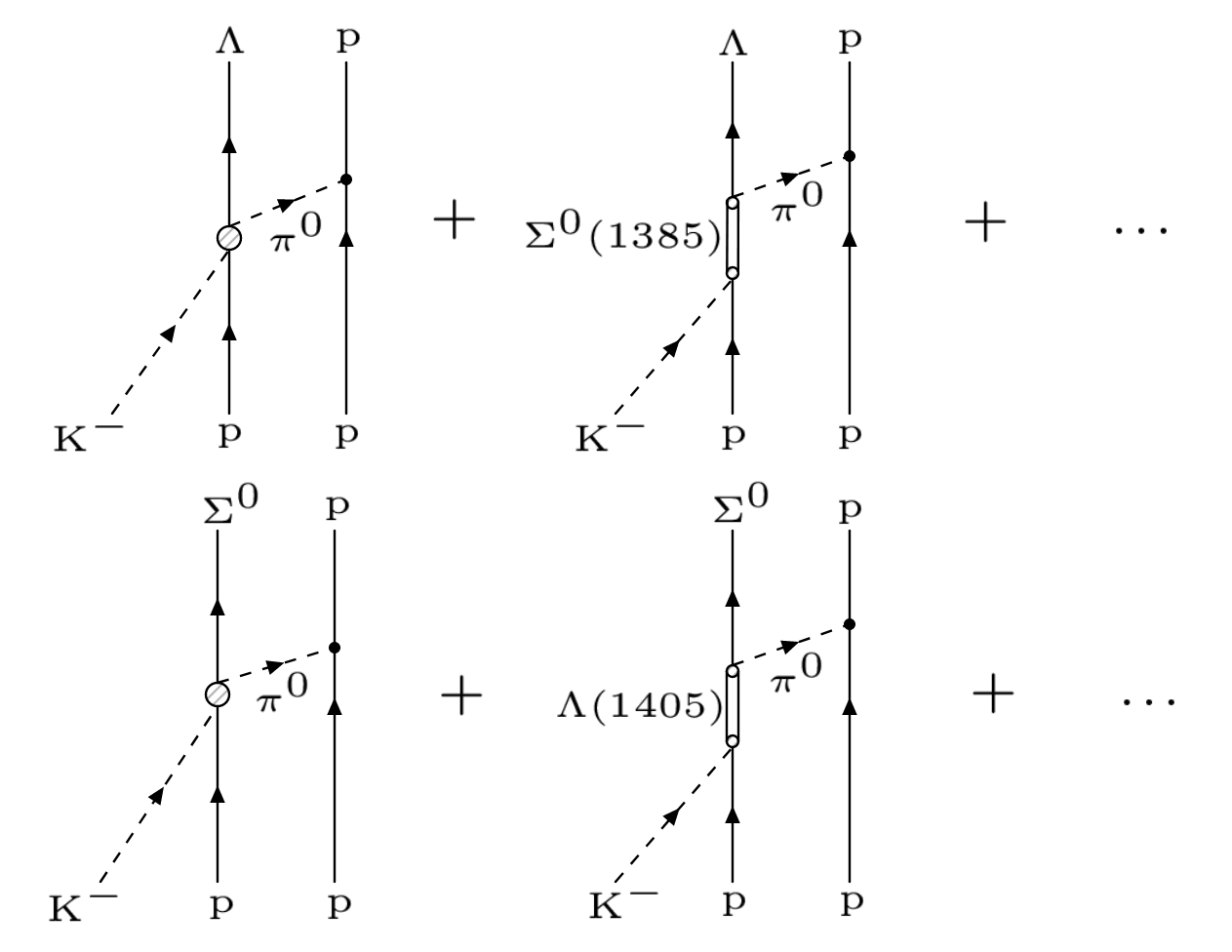}
\caption{The tree level (left) and the leading order (right) diagrams of the processes $\mathrm{K^-(pp) \rightarrow \Lambda p}$ (top) and $\mathrm{K^-(pp) \rightarrow \Sigma^0 p}$ (bottom), according to the meson exchange model.}
\label{fdiagrams}
\end{figure}

Considering the meson exchange model, the processes $\mathrm{K^-(pp) \rightarrow \Lambda p}$ and $\mathrm{K^-(pp) \rightarrow \Sigma^0 p}$ occur with the exchange of a $\pi^0$ (see Ref. \cite{OT2006}) according to the diagrams shown in Fig. \ref{fdiagrams}. 

As a consequence, the ratio of the BRs of the 2NA-QF \Lambdan p and the 2NA-QF \Sigman$^0$p is directly related to the ratio of BRs of the $\mathrm{K^- (p) \rightarrow \Lambda \pi^0}$ and the $\mathrm{K^- (p) \rightarrow \Sigma^0 \pi^0}$ processes:
\begin{equation}
\mathrm{\mathcal{R} = \frac{BR (K^-(pp) \rightarrow \Lambda p)}{BR (K^-(pp) \rightarrow \Sigma^0 p)} = \frac{BR(K^-(p) \rightarrow \Lambda \pi^0)}{BR(K^-(p) \rightarrow \Sigma^0 \pi^0)} \ .}
\label{ratioBR}
\end{equation}
The exchange of a K$^-$ meson is also possible \cite{wycech1967,Sekihara2009,Sekihara2012}. Detailed calculations dating from the emulsion studies period show that the effect of the latter is a 30\% correction to the pion exchange \cite{wycech1967}. Nevertheless, the two mechanisms add coherently and equation (\ref{ratioBR}) still holds.   
Given that the \Lambdan$  \pi^0$ and \Sigman$^0 \pi^0$ production below the \kbarN threshold involve the intermediate formation of the \Sigman$^{0}$(1385) and \Lambdan(1405) resonances, the measured ratio gives information on the complex and debated properties of these resonances.

Using the branching ratios in Table \ref{BRandCS} we measure:
\begin{equation}
\mathrm{
\mathcal{R} = 0.7 \pm 0.2 (stat.) {}^{+0.2}_{-0.3} (syst.) \ .
}
\label{ratiolps0p}
\end{equation} 

This ratio is to be compared with the ratio $\mathcal{R}'$ between the phase spaces of the two reactions:
\begin{equation}
\mathrm{
K^- + {}^{12}C \rightarrow K^- + (pp) + R \rightarrow \Lambda + p + R 
}
\nonumber
\end{equation} 
and 
\begin{equation}
\mathrm{
K^- + {}^{12}C \rightarrow K^- + (pp) + R' \rightarrow \Sigma^0 + p + R' \ , 
} \nonumber
\end{equation}
which is given by:
\begin{equation}
\mathrm{\mathcal{R}' = \frac{\sum_{R} \cdot w_R \cdot \int |F(\textit{p}_{\Lambda p})|^2 \cdot d\rho_{\Lambda p} }{\sum_{R'} \cdot w_{R'} \cdot \int |F(\textit{p}_{\Sigma^0 p})|^2 \cdot d\rho_{\Sigma^0 p}} = 1.22 \ .}
\label{ratio}
\end{equation}
The sum is performed over residual nuclei combinations considered in the fit, where w$_\mathrm{R}$ and w$_\mathrm{R'}$ are the relative weights extracted from the fit. 

By comparing (\ref{ratio}) with (\ref{ratiolps0p}) we see that the ratio of the pure phase spaces ($\mathcal{R}^\prime$) is bigger then $\mathcal{R}$ which contains also the dynamical effects besides the phase spaces. This indicates that the I=0 \Lambdan(1405) production cross section, contributing to the denominator of $\mathcal{R}$, is dominant with respect to the I=1 \Sigman$^0$(1385) production cross section, involved in the numerator of $\mathcal{R}$.
This measurement provides crucial information on the \Lambdan(1405) in medium properties and on the resonance couplings to the \kbarN and \Sigman$\pi$ channels.
\subsection{Search for a K$^-$pp Bound State}
In the following, a contribution of an eventual K$^-$pp bound state to the measured \Lambdan p distributions is investigated. The K$^-$pp bound state formation in Carbon
\begin{equation}
\mathrm{
K^- + {}^{12}C \rightarrow K^-pp + {}^{10}Be \rightarrow \Lambda + p + {}^{10}Be \ ,
} \nonumber
\end{equation}
followed by the K$^-$pp decay in \Lambdan p, was simulated according to the procedure described in Section \ref{sec:MC}, using a Breit-Wigner shape for $\mathrm{|T(\mathit{m}_{\Lambda p})|^2}$.

In order to test the sensitivity of the experimental spectra to an eventual K$^-$pp signal, the kinematic shapes of the K$^-$(pp) 2NA process are compared with bound states simulations, selecting for the bound state values the binding energies and widths which are claimed in previous experimental findings: E15 (first run), FINUDA (which exploit K$^-$ beams), E27 ($\pi$ induced reactions), DISTO (p-p collisions) and OBELIX (antiproton annihilation on $^4$He), listed in Table \ref{tabexp}. In the case of OBELIX the upper limit of the bound state width is used.
\begin{table*}[!h]
\centering
\caption{Experimental BE and $\Gamma$ of the K$^-$pp bound state measured by the FINUDA, OBELIX, DISTO, E27 and E15 (first run) experiments.}
\resizebox{0.9\linewidth}{!}{
\begin{tabular}{ccc}
\hline
Experiment & BE (MeV) & $\Gamma$ (MeV/c$^2$) \\ [0.5ex]
\hline\vspace{0.1cm}
FINUDA \cite{FINUDA2005} & 115 $^{+6}_{-5}$ \small{(stat.)} $^{+3}_{-4}$ \small{(syst.)} & 67 $^{+14}_{-11}$ \small{(stat.)} $^{+2}_{-3}$ \small{(syst.)} \\[0.5ex]
OBELIX \cite{OBELIX} & 160.9 $\pm$ 4.9 & $<$ 24.4 $\pm$ 8.0  \\[0.5ex]
DISTO \cite{DISTO} & 103 $\pm$ 3 \small{(stat.)} $\pm$ 5 \small{(syst.)} & 118 $\pm$ 8 \small{(stat.)} $\pm$ 10 \small{(syst.)} \\[0.5ex]
E27 \cite{E27} & 95 $^{+18}_{-17}$ \small{(stat.)} $^{+30}_{-21}$ \small{(syst.)} & 162 $^{+87}_{-45}$ \small{(stat.)} $^{+66}_{-78}$ \small{(syst.)}  \\[0.5ex]
E15 (first run) \cite{E15} & 15 $^{+6}_{-8}$ \small{(stat.)} $\pm$ 12 \small{(syst.)} & 110 $^{+19}_{-17}$ \small{(stat.)} $\pm$ 27 \small{(syst.)} \\[0.5ex]
\hline
\end{tabular}
}
\label{tabexp}
\end{table*}

\begin{figure*}[!h]
\centering
\includegraphics[width=0.7\textwidth]{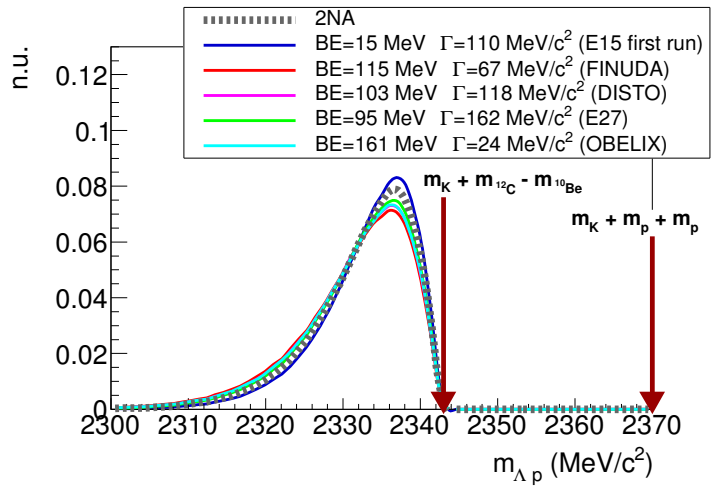}
\caption{Calculated invariant mass distributions of the \Lambdan p pairs from the K$^-$pp bound state decay, produced in K$^-$ absorption on $^{12}$C. The line shape of the 2NA (gray line) is shown for comparison together with the shapes obtained for the simulation of a bound state having binding energy and width selected according to the measurements of E15 first run (blue line), FINUDA (red line), DISTO (magenta line), E27 (green line) and OBELIX (cyan line). The areas of the distributions are normalised to unity.}
\label{bsmeasure}
\end{figure*}

The calculated \Lambdan p invariant mass distributions, corresponding to the parameters listed in Table \ref{tabexp}, are shown in Fig. \ref{bsmeasure}, together with the shapes for the K$^-$(pp) 2NA absorption process.

The comparative study evidences that, for low-energy K$^-$ absorption in $^{12}$C, a bound state characterised by the quoted parameters can not be distinguished from the 2NA.

\begin{figure*}[!h]
\centering
\includegraphics[width=0.7\textwidth]{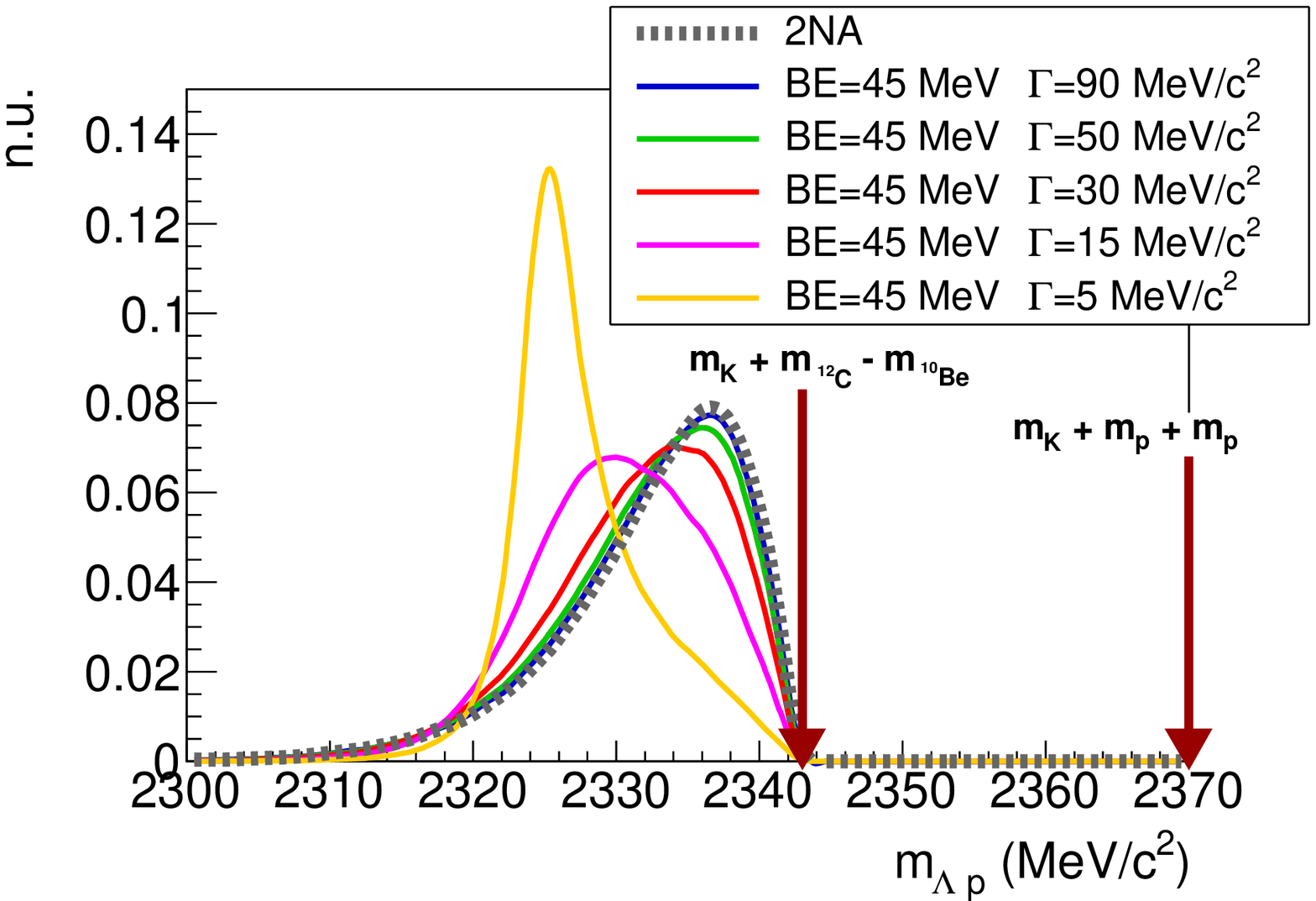}
\caption{Calculated \Lambdan p invariant mass distributions for the process $\mathrm{K^- + {}^{12}C \rightarrow K^-pp + {}^{10}Be \rightarrow \Lambda + p + {}^{10}Be}$, for a bound state having BE = 45 MeV and $\Gamma$ = 5, 15, 30, 50 and 90 MeV/c$^2$ (yellow, magenta, red, green and blue curves respectively). The gray curve is the shape of the 2NA-QF. The areas of the distributions are normalised to unity.}
\label{bs45g}
\end{figure*}

In order to further test the experimental sensitivity to the width of the bound state, the same kinematic distributions where calculated for a fixed binding energy (BE=45 MeV) and various widths ($\Gamma$ = 5, 15, 30, 50 and 90 MeV/c$^2$). The invariant mass shapes are shown in Fig. \ref{bs45g}. It emerges that a bound state signal could be disentangled by the 2NA only for extremely narrow states ($\Gamma < 15$ MeV/c$^2$), excluded by theoretical predictions \cite{AY2002,IS2007,SGM2007,WG2009,Revai,MAY2013,DHW2009,BGL2012,IKS2010,Bicudo,BO2013}. The same result is obtained also for other choices of the BE.

From Table \ref{BRandCS}, the total branching ratio of the 2NA process decaying into \Lambdan p (QF and with FSIs) is:
\begin{equation}
\mathrm{
BR(K^- 2NA \rightarrow \Lambda p) = ( 6.5 \pm 1.2 (stat.) {}^{+0.5}_{-0.6} (syst.) ) \% \ ,
}
\nonumber
\end{equation} 
and it contains the K$^-$pp contribution, if such a state exists.
\subsection{Comparison with the FINUDA result}\label{sec:FINUDA}
\begin{figure}[!h]
\centering
\includegraphics[width=0.7\textwidth]{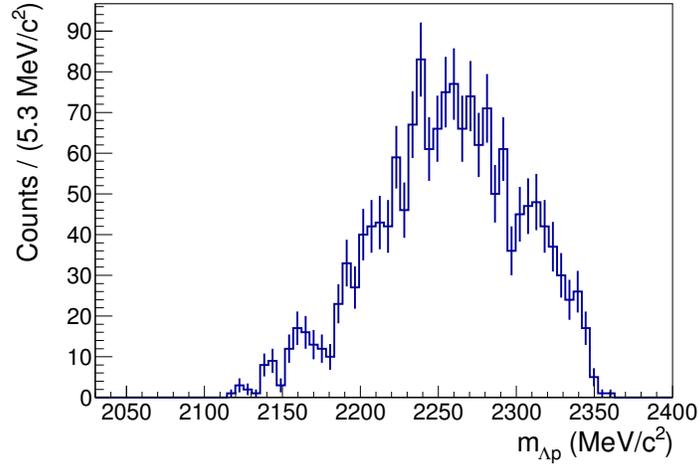}
\caption{The experimental \Lambdan p invariant mass spectrum obtained selecting back-to-back \Lambdan p events ($\cos \theta_\mathrm{\Lambda p} < -0.8$), in analogy with the data analysis performed by FINUDA in Ref \cite{FINUDA2005}.}
\label{LPminvFINUDA}
\end{figure}

It is interesting to compare our \Lambdan p invariant mass spectrum with the one measured by FINUDA \cite{FINUDA2005}, which is interpreted as a K$^-$pp bound state decay. The target used in Ref.~\cite{FINUDA2005} is a mixture of ${}^{12}$C, ${}^{6}$Li and ${}^{7}$Li, in the following proportions: 51 \%, 35 \% and 14 \% respectively. In Ref. \cite{OsetFINUDA} is, however, demonstrated that the $^{12}$C component dominates. This is motivated by the larger amount of protons, the higher probability of the FSI in the bigger nucleus and by the larger overlap with the K$^-$ wave function. In order to perform a direct comparison we included in our selection the same angular distribution cut which was performed in Ref. \cite{FINUDA2005} to pin down the back-to-back \Lambdan p production, which is argued in Ref. \cite{FINUDA2005} to be directly correlated to the K$^-$pp bound state production:
\begin{equation}
\cos \theta_\mathrm{\Lambda p} < -0.8 \ .
\end{equation}
The obtained \Lambdan p invariant mass spectrum is shown in Fig. \ref{LPminvFINUDA}. 

\begin{figure*}[!h]
\centering
\includegraphics[width=0.49\textwidth]{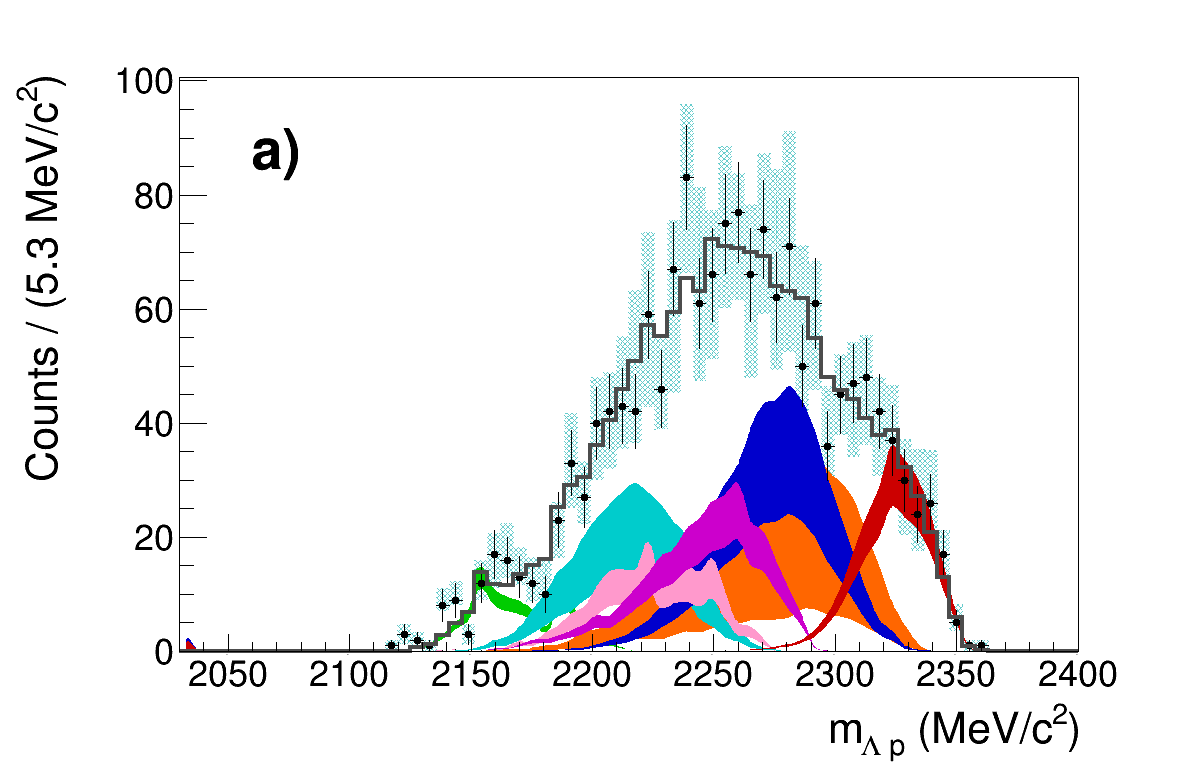}
\includegraphics[width=0.49\textwidth]{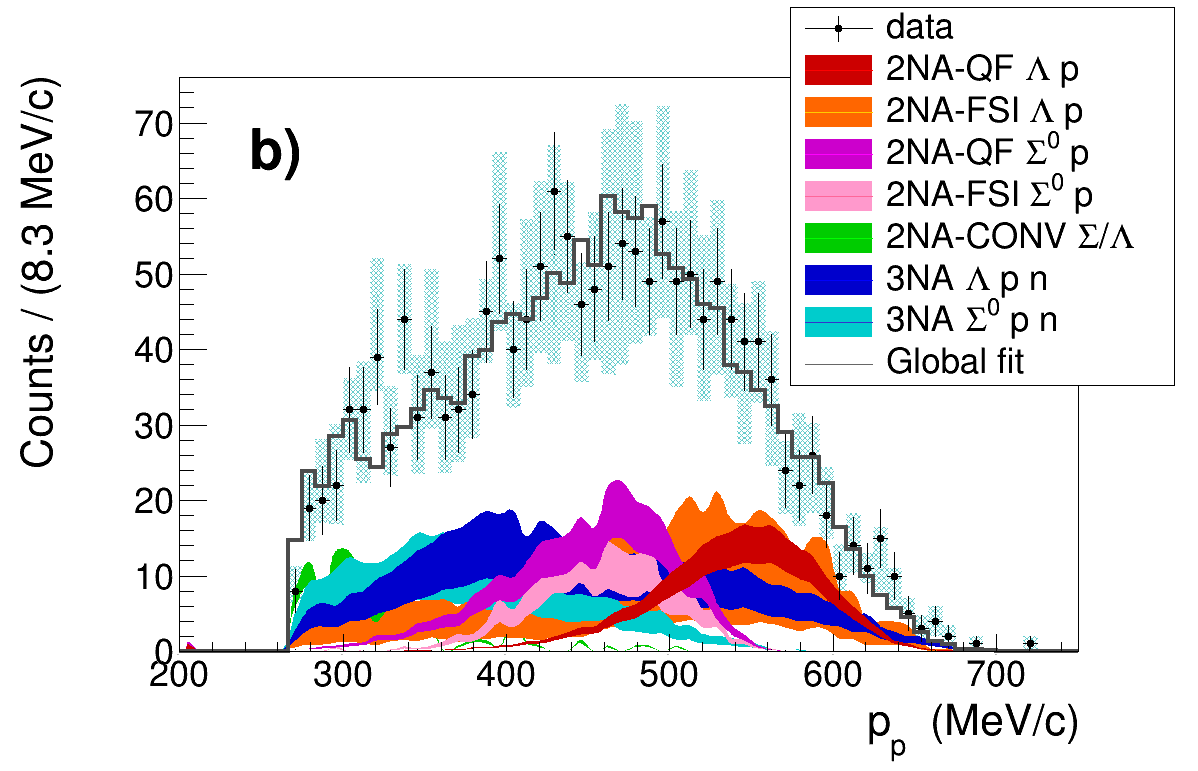}
\caption{
\Lambdan p invariant mass (panel a) and proton momentum (panel b) spectra. Black points represent the data, black error bars correspond to the statistical errors, cyan error bars correspond to the systematic errors. The gray line distributions represent the global fitting functions. The color legend for the different contributing processes is shown in the figure. }
\label{LPallFINUDA}
\end{figure*}

The shape of the \Lambdan p invariant mass spectrum in Fig. \ref{LPminvFINUDA} is compatible with the corresponding spectrum reported by the FINUDA collaboration in Fig. 3 of Ref.~\cite{FINUDA2005}.
Performing the simultaneous fit of the \Lambdan p invariant mass, the \Lambdan p angular correlation, the \Lambdan ~ and the proton momenta, in complete analogy with the procedure described for the fit of our complete selected \Lambdan p data sample, a reduced $\chi^2$ of $\chi^2/ndf = 123/130 = 0.94$ is obtained. The result of the fit is shown in detail for the invariant mass and proton momentum components in Fig. \ref{LPallFINUDA}. 

\begin{table}
\centering
\caption{The branching ratios of K$^-$ multi-nucleon absorptions for a back-to-back (a la FINUDA) selection of \Lambdan ~and p are reported together with the statistical and systematic errors.}
\resizebox{0.7\linewidth}{!}{
\begin{tabular}{ll}
\hline
Process & Branching Ratio (\%)\\ [0.5ex]
\hline\vspace{0.1cm}
2NA-QF \Lambdan p & 0.20 $\pm$ 0.04(stat.) $\pm$ 0.02(syst.) \\[0.5ex]
2NA-FSI \Lambdan p & 3.8 $\pm$ 2.3(stat.) $\pm$ 1.1(syst.) \\[0.5ex]
2NA-QF \Sigman$^0$p & 0.54 $\pm$ 0.20(stat.) ${}^{+0.20}_{-0.16}$(syst.) \\[0.5ex]
2NA-FSI \Sigman$^0$p & 5.4 $\pm$ 1.5(stat.) ${}^{+1.0}_{-2.7}$(syst.) \\[0.5ex]
2NA-CONV \Sigman /\Lambdan & 22 $\pm$ 4(stat.) ${}^{+1}_{-12}$(syst.) \\[0.5ex]
3NA \Lambdan pn & 1.1 $\pm$ 0.3(stat.) $\pm$ 0.2(syst.) \\[0.5ex]
3NA \Sigman$^0$pn & 1.9 $\pm$ 0.7(stat.) ${}^{+0.8}_{-0.4}$(syst.) \\[0.5ex]
\hline
\end{tabular}
}
\label{brancing ratiosFINUDA}
\end{table}

The BRs per stopped K$^-$ of the various contributing processes are again evaluated using the parameters obtained from the fit of the back-to-back \Lambdan p events and are presented in Table \ref{brancing ratiosFINUDA}. The BRs are compatible with those obtained from the fit of the full selected data sample reported in Table \ref{BRandCS}, in both cases the measured distributions can be well explained in terms of K$^-$ multi-nucleon absorption processes. The fit to the back-to-back \Lambdan p sample is not sensitive to the contribution of the \Sigman/\Lambdan~ conversion process which is the most critically affected by the additional phase space cut as evidenced by the big systematic error.  

Concluding, even imposing the back-to-back angular selection to pin down a highly correlated \Lambdan p production, the spectrum can be completely explained in terms of K$^-$ multi-nucleon absorption processes, without the need of a K$^-$pp component, whose partial contribution, if any, is contained in the 2NA process.

\section{Summary}
In this work the correlated \Lambdan p pairs production from low-energy K$^-$ $^{12}$C absorption was investigated, taking advantage of the negatively charged kaons produced by $\phi$s decaying at-rest at the DA\Phin NE collider. A sample of almost pure K$^-$ $^{12}$C interactions is obtained by selecting the K$^-$ captures in the inner wall of the KLOE drift chamber, which is used as an active target. The key selection criterion which characterises the \Lambdan p correlated pairs production search, followed in this work, is represented by the measurement of the mass by time of flight for both the involved protons. One proton is originated in the $\mathrm{\Lambda \rightarrow p \pi^-}$ decay, one is produced at the K$^-$ multi-nucleon absorption. At the price of a reduced statistics this selection returns an extremely clean data sample. The acceptance cut which results from the protons detection in the calorimeter (necessary for the time of flight measurement) suppresses the K$^-$ single nucleon absorption (1NA) contributions. The calorimeter momentum threshold ($p > 240$ MeV/c) is indeed higher than the typical momentum of protons emerging from pionic productions. The residual 1NA contamination, surviving the selection cuts, is considered as a contribution to the systematic errors.

BRs and cross sections of the K$^-$ multi-nucleon absorptions on two, three and four nucleons (2NA, 3NA and 4NA) were obtained by means of a simultaneous fit of the \Lambdan p invariant mass, \Lambdan p angular correlation, \Lambdan ~ and proton momenta to the simulated distributions for both direct \Lambdan ~ production and \Sigman$^0$ production followed by $\mathrm{\Sigma^0 \rightarrow \Lambda \gamma}$ decay. The K$^-$ nuclear capture was calculated for both at-rest and in-flight interactions. In the first case the absorption from atomic 2p state is assumed. Fragmentations of the residual nucleus following the hadronic interaction were also considered. For the 2NA the important contributions of both final state interactions (FSI) of the \Lambdan ~ and the proton were taken into account, as well as the conversion of primary produced sigma particles ($\mathrm{\Sigma N \rightarrow \Lambda N'}$); this allows to disentangle the quasi-free (QF) production. The global branching ratio for the K$^-$ multi-nucleon absorption in $^{12}$C (with \Lambdan (\Sigman$^0$)p final states) is found to be compatible with bubble chamber results. The measured BRs and low-energy cross sections of the distinct K$^-$ 2NA, 3NA and 4NA will be useful for the improvement of microscopical models of the K$^-$NN absorption and for future generalisation to K$^-$ absorption reaction calculations involving even more than two nucleons. 

The \Lambdan p direct production in 2NA-QF is phase space favoured with respect to the corresponding \Sigman$^0$p final state, the ratio between the final state phase spaces for the two processes is $\mathcal{R}' \simeq 1.22$. The measured ratio for the BRs of the two processes is instead $\mathcal{R} = \mathrm{0.7 \pm 0.2 (stat.) {}^{+0.2}_{-0.3} (syst.)}$, indicating that the \Sigman$^0$p production dominates over  \Lambdan p production. This can be interpreted as a consequence of the bigger \Lambdan(1405) production cross section (contributing to the \Sigman$^0$p channel) with respect to the \Sigman$^0$(1385) production cross section (involved in the  \Lambdan p channel). From $\mathcal{R}$ and $\mathcal{R}^\prime$ important information can be extracted on the $\bar{\mathrm{K}}$N dynamics below threshold.

The last part of this work is devoted to the search for K$^-$pp bound state contribution to the measured \Lambdan p production.
The simulated distributions were obtained in agreement with the multi-nucleon K$^-$ capture calculation assuming a Breit-Wigner distribution for the bound state formation transition amplitude. The 2NA-QF is found to completely overlap with the K$^-$pp, except for small values of the bound state width of the order of 15 MeV/c$^2$ or less. 
A further selection of back-to-back \Lambdan p production was performed by selecting $\cos\theta_\mathrm{\Lambda p} < -0.8$ in order to make a direct comparison with the corresponding FINUDA measurement. The shape of the invariant mass distribution is perfectly compatible with the shape presented in Ref. \cite{FINUDA2005}. By repeating the fitting procedure the spectra result to be completely described in terms of K$^-$ multi-nucleon absorption processes, the BRs are compatible with those obtained from the fit of the full data sample, with no need of a K$^-$pp component, whose partial contribution, if any, is contained in the 2NA.

\section*{Acknowledgements}
We acknowledge the KLOE/KLOE-2 Collaboration for their support and for having provided us the data and the tools to perform the analysis presented in this paper. 
Part of this work was supported by Minstero degli Affari Esteri e della Cooperazione Internazionale, Direzione Generale per la Promozione del Sistema Paese (MAECI), Strange Matter project PRG00892;
Polish National Science Center through grant No. UMO-2016/21/D/ST2/01155.

\bibliography{elsarticle-template-num}






\end{document}